\documentclass[preprint, 12pt, numafflabel]{elsarticle}

\usepackage{array}
\usepackage[title, titletoc]{appendix}
\setcitestyle{square}
\usepackage{adjustbox}
\usepackage{tabularx}
\usepackage{afterpage}

\makeatletter
\g@addto@macro\@floatboxreset\centering

\makeatother
\usepackage{fancyhdr}
\usepackage[utf8]{inputenc}
\usepackage{graphicx} 
\usepackage{amsmath}
\usepackage{subcaption}
\usepackage{upgreek}
\usepackage{booktabs}
\usepackage{multirow}

\usepackage{pdflscape}
\usepackage{rotating}

\usepackage{hyperref}
\hypersetup{colorlinks,allcolors=black}
\usepackage[top=2cm, bottom=2cm, left=2cm, right=2cm]{geometry}

\newcommand{\exvivo}{\textit{ex vivo}}
\newcommand{\Exvivo}{\textit{Ex vivo}}
\newcommand{\invivo}{\textit{in vivo}}
\newcommand{\Invivo}{\textit{In vivo}}
\newcommand{\includepageasasubfigure}[1]{\begin{subfigure}{\linewidth}
                \includegraphics[width=\textwidth,page=#1]{images/ALL_SAMPLES.pdf}
            \end{subfigure}}      
\usepackage{amssymb}

\usepackage{lineno}

\date{February 2025}

\begin{document}
    \journal{A définir}
    \begin{frontmatter}
    
        \title{Refraction corrected specular beamforming applied to cortical bone enhances interface visibility of bone-soft tissues interfaces}
        
        \affiliation[lib]{{organization}={Sorbonne Université, INSERM, CNRS, Laboratoire d’Imagerie Biomédicale, LIB, F-75006},
                    city={Paris},
                    country={France}}
        \affiliation[delft]{organization={Department of Imaging Physics, Delft University of Technology},
                    country={The Netherlands}}
        
        \author[lib]{Amadou S. DIA \texorpdfstring{\footnote{Corrresponding author: amadou.dia@sorbonne-universite.fr}{}}}
        
        \author[lib]{Quentin GRIMAL}
        \author[lib,delft]{Guillaume RENAUD}
    \end{frontmatter}
\section*{Abstract}
Ultrasound imaging of the cortex of long bones may enable the measurement of the cortical thickness and the ultrasound wave speed in cortical bone tissue. However, with bone loss, the cortical porosity and the size of the vascular pores increase, resulting in strong ultrasound diffuse scattering whose magnitude can exceed that of the specular reflection from the bone cortex-marrow (endosteal) interface.

In this study we adapt to bone a specular beamforming technique proposed to better image a needle in soft tissue. Our approach takes into account both wave refraction and specular reflection physics to enhance the contrast of bone surfaces and reduce speckle from intracortical pores.

In vivo ultrasound data were acquired at the center of the human tibia in a plane normal to the bone axis. For \exvivo~measurements, we analyzed 16 regions of interest from the femoral diaphysis of three elderly donors (donors 66-98 y.o.) using a 2.5 MHz US transducer. A single-element transmission synthetic aperture imaging sequence was implemented on a research ultrasound system with a 2.5MHz phased array transducer. Image reconstruction was performed using two different reconstruction methods: (A) a delay-and-sum (DAS) algorithm with optimized f-number, correction of refraction at the soft tissue-bone interface and subject-specific ultrasound wave speed and (B) an adaptive algorithm using Snell’s law of reflection. The improvement of image quality was evaluated with contrast ratios of the average intensities: C$_{EI}$ between the endosteal surface and the center of the cortex.

\Invivo, specular beamforming improved the visibility of the endosteum (C$_{EI}$) by 1 to 13 dB while maintaining the relative contrast between the outer and inner surfaces of the cortex. These results suggest that the visualization of the intra-osseous anatomy can be enhanced if Snell’s law and wave refraction are taken into account during image reconstruction.
\clearpage
\section{Introduction}
\begin{figure}[htb!]
        \begin{subfigure}{.65\textwidth}
            \includegraphics[width=\textwidth]{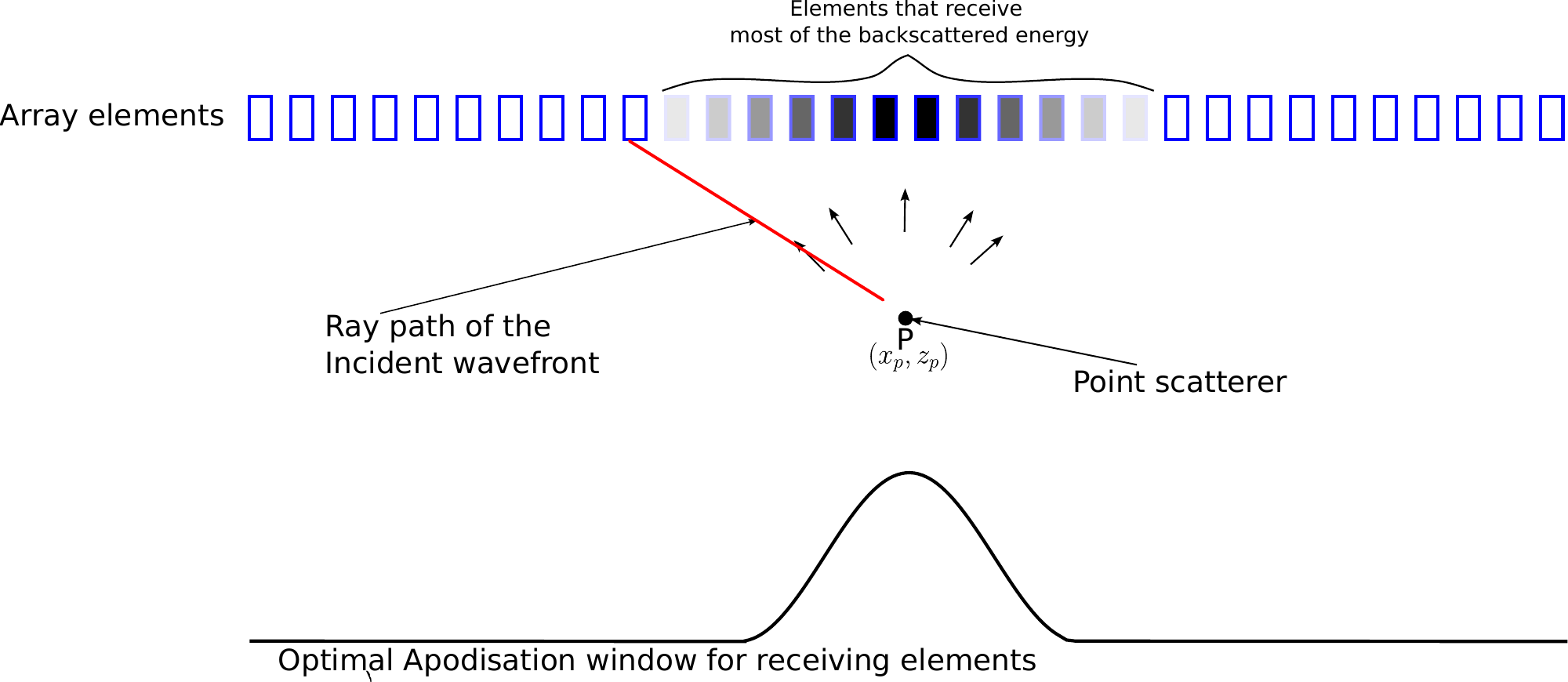}
            \caption{Diffuse scatterer.}
            \label{chap5:fig1:subfiga:difuse}
        \end{subfigure}
      \begin{subfigure}{.65\textwidth}
            \includegraphics[width=\textwidth]{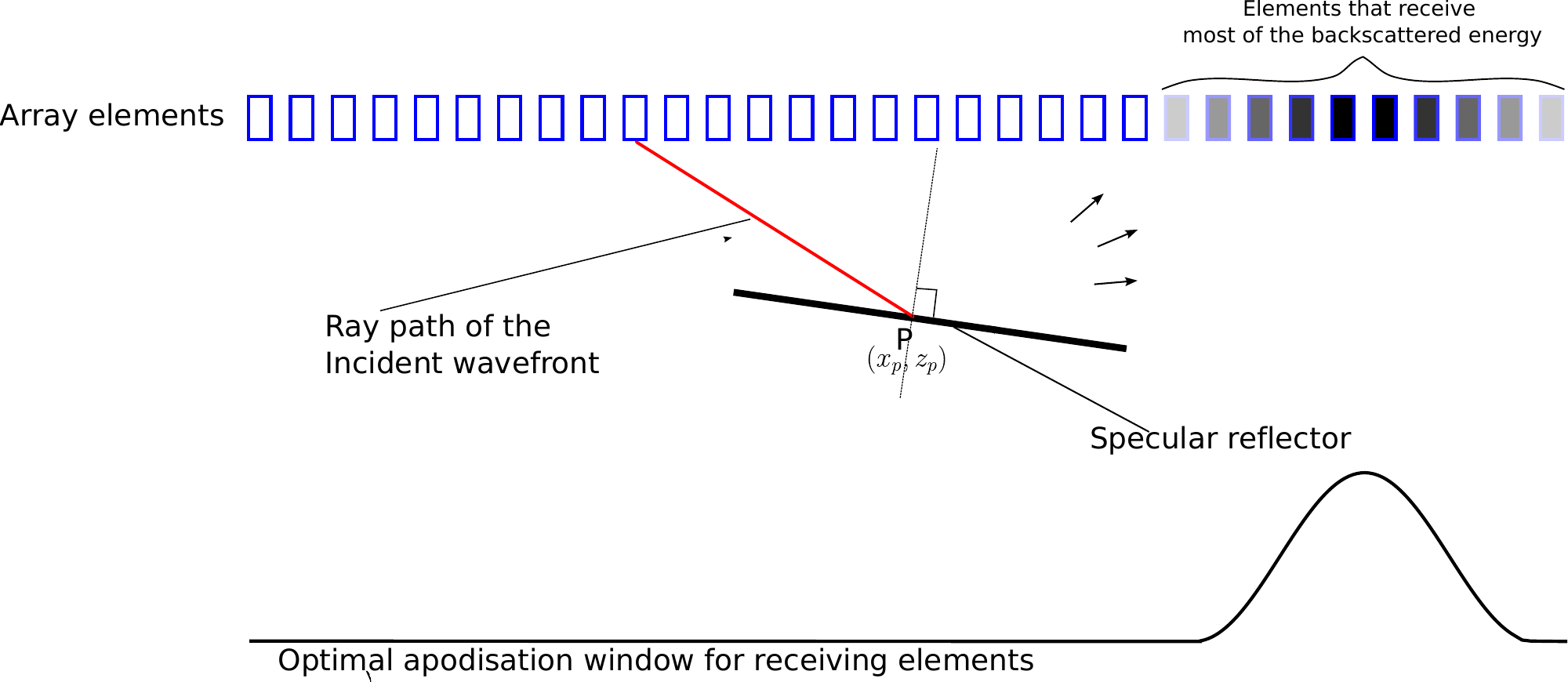}
            \caption{Specular scatterer.}
            \label{chap5:fig1:subfigb:specular}
    \end{subfigure}
        \caption{Illustration of the optimal receive sub-aperture for the reconstruction of pixel $P$. In (a) maximum backscattered amplitude is recorded by the nearest element of the transducer array. In (a) most of the back-scattered energy is concentrated to receivers around the lateral position of the scatterer. In (b), maximum backscattered amplitude is not recorded by the nearest element of the probe array.}
        \label{chap5:fig1:diffuse_vs_specular_apod}
    \end{figure}

    In medical ultrasound imaging, DELAY-AND-SUM (DAS) \cite{perrot_so_2021} is the most widely used beamformer for image reconstruction. Beamformers based on DAS assume a homogeneous medium composed only of point scatterers (small compared to the wavelength). For a diffusive point scatterer, the signal is backscattered uniformly in all directions. Therefore, maximum backscattered amplitude is recorded by the nearest element of the transducer array, because of diffraction loss (see Figure~\ref{chap5:fig1:subfiga:difuse}). In a typically DAS implementation, the receive sub-aperture is chosen according to geometry and element directivity: it is independent of data.
    
    DAS beamforming can yield good image quality for a nearly planar specular interface (much wider than wavelength) and nearly parallel to the surface of the ultrasound probe. However, when the interface is not parallel to the probe surface, maximum backscattered amplitude is not recorded by the nearest element of the probe array. The tilting angle of the interface determines the main direction of the reflected wavefront (Snell-Descartes' law). (see Figure~\ref{chap5:fig1:subfigb:specular}).

    At 2.5~MHz (frequency previously used for \invivo~measurements \cite{renaud_vivo_2018,salles_revealing_2021}), there exists both specular and diffuse reflection of ultrasound inside cortical bone. In fact, on the external (periosteal) and internal (endosteal) surfaces of the cortex, the large impedance ratio between soft tissue and cortical bone tissue leads to specular reflection and cortical pores act as scatterers smaller than the wavelength. Therefore, we expect a significant effect of specular reflection in bone ultrasound images.

    As discussed in \cite{dia_influence_2023, dia_ultrasound_2025}, achieving a high-quality image of the cortex is crucial for clear visualization of the periosteal and endosteal surfaces. Specifically, \cite{dia_influence_2023, dia_ultrasound_2025} highlighted that during DAS beamforming, diffuse signals from the microstructure create speckle that may obscure the specular signal from the endosteal surface of the bone. An adaptive reconstruction technique could enhance signals from both the external and internal surfaces of cortical bone. We hypothesize that an optimal image reconstruction of bone cortex interfaces could be obtained with an adaptive receive sub-aperture determined based on the properties of specular reflection.

    In this study, we present the principle of a beamforming algorithm tailored to the physics of specular reflection and refraction, with the specific goal of improving the visibility of the external and internal interfaces of the bone cortex, for instance to measure the cortical thickness.

    In recent years, various approaches have been proposed for extracting specular information in medical ultrasound imaging. In 2008,\textit{ Vogt et al} \cite{vogt_parametric_2008} suggested using a single element transducer and exploiting the laws and properties of specular reflections, to reconstruct parametric images. Ultrasound data are acquired from different angular tilt and lateral positions of the transducer. The parametric images contain first-order statistics such as mean, standard deviation, maximum, and minimum of the envelope of received echo signals across all emission-reception events of specular reflections. By analyzing the parametric images, specular reflection could be effectively differentiated from diffuse scattering.  
    
    Similarly, \textit{Bandaru et al} \cite{bandaru_delay_2016} proposed using an array transducer to enhance reflections from specular interfaces. Instead of using the average, as in DAS methods, they took the standard deviation of received backscattered signals across the receive aperture. The orientation of the specular interface was estimated based on the peak echo amplitude. This method was developed for conventional focusing in transmit.
    
    In a preliminary study, \textit{Nagaoka et al}~\cite{nagaoka_preliminary_2020} suggested the use of a data-independent apodization weight to highlight both diffuse scattering from small heterogeneities and specular reflection from flat interfaces.
    
    \textit{Rodriguez-Molares et al} \cite{rodriguez-molares_specular_2017} proposed a physical model-based technique of emphasizing specular reflection for needle tracking using a synthetic aperture sequence. Their technique, based on the source-image principle, involves developing a specular reflection model. They compared this model with the coherent compounding of received signals that follow Snell's law to obtain a matched filter maximizing the signal-to-noise ratio (SNR) of signals reflected by planar interfaces. This technique suppresses speckle and enhances the visualization of specular interfaces, such as an inserted needle.
    
    More recently, \textit{Malamal et al} \cite{malamal_enhanced_2023} proposed an innovative approach involving the radon transform and plane wave imaging to identify the receiver index that maximizes the back-scattered energy from a specular interface. Then, a receive apodization window, centered around this optimized receiver index, is selected during Delay-and-Sum (DAS) beamforming. The same group \cite{malamal_physics_2023} proposed to provide a real-time visual feedback mechanism for operators. They introduced a color-coded image containing a vectorized estimation of the reflection directivity within a defined region of interest. This allows the operator to reorient the probe or adjust the transmission sequence to align parallel to the surface of the reflector. Employing a plane wave imaging sequence, they applied this method to synthetic and experimental data for needle tracking and external bone surface imaging, effectively distinguishing fractured and smoothed regions of the periosteal surface of a bone. However, like conventional DAS algorithms, all these methods assume a homogeneous medium and a straight ray-path propagation hypothesis. 
    
    In this study, we implement specular beamforming according to \cite{rodriguez-molares_specular_2017} adapted to bone imaging. To emphasize the reflection of the external and internal surfaces of the bone, we propose to consider the physics of reflection and refraction. 


\section{Refraction-corrected specular beamforming}
\subsection{Geometrical considerations}
\begin{figure}[htb!]
    \includegraphics[width=.7\textwidth]{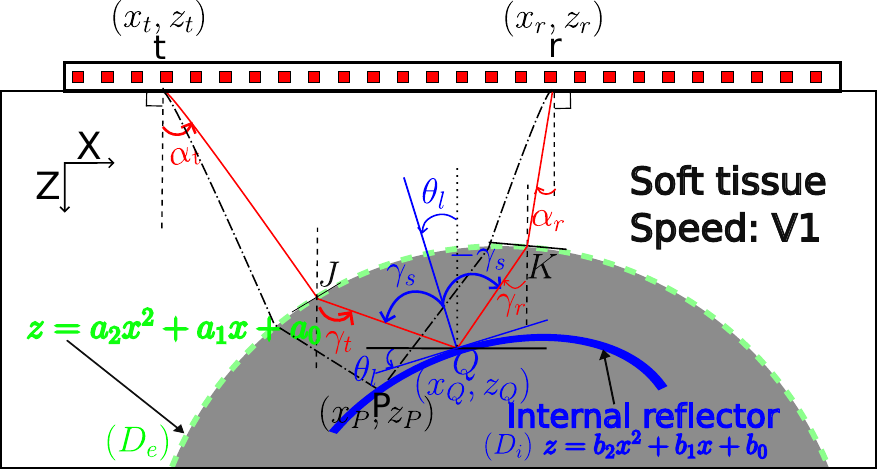}
    \caption{Illustration of specular reflection inside cortical bone.}
    \label{fig:refraction_reflection}
\end{figure}

We suppose a medium with two layers: a layer of cutaneous tissue on top of  a layer of bone tissue. The interface between the two layers can be approximated by a parabola with parameters $a_0, a_1, a_2$ :
\begin{equation}
    \label{chap5:eq:periosteum_equation}
    (D_e): z=a_0+a_1x+a_2x^2,
\end{equation}
where x and z are along the axes defined in Figure~\ref{fig:refraction_reflection},$V_1$ and $V_2$ are the propagating wave speed inside cutaneous tissue and cortical bone respectively. We define $\alpha_t$ (respectively $\alpha_r$) as the angle of the incident ray parting (respectively the reflected ray) from element $iT$ (respectively received at element $iR$) of the probe and $\gamma_t$ (respectively $\gamma_r$) as the angle of the ray arriving at (respectively parting from) point $P$ inside cortical bone (see Figure~\ref{fig:refraction_reflection}). 

We define incident point $J$ as the point of the interface through which the incident ray passes and incident point $K$ as the point of the interface through which the reflected ray passes. Using these points above angles are given by:
\begin{equation}
    \label{chap5:eq:angle_ref}
    \left\{
    \begin{array}{ll}
        \alpha_t = \arctan(\frac{x_J-x_t}{z_J-z_t})\,\\
        \gamma_t = \arctan(\frac{x_P-x_J}{z_P-z_J})\,\\
        \alpha_r = \arctan(\frac{x_K-x_r}{z_K-z_r})\,\\
        \gamma_r = \arctan(\frac{x_P-x_K}{z_P-z_K})
    \end{array}
    \right.
\end{equation}

According to Snell-Descartes law of refraction for the transmitted ray from $iT$ to $P$:
\begin{equation}
    \label{chap5:eq:refraction_transmit}
    \frac{\sin(\alpha_t(P)+\arctan(2a_2x_J+a_1))}{V_1} = \frac{\sin(\gamma_t(P)+\arctan(2a_2x_J+a_1))}{V_2}.
\end{equation}
        
The reflected ray from $P$ to $iR$ follows the same laws:
\begin{equation}
    \label{chap5:eq:refraction_receive}
    \frac{\sin(\alpha_r(P)+\arctan(2a_2x_K+a_1))}{V_1} = \frac{\sin(\gamma_r(P)+\arctan(2a_2x_K+a_1))}{V_2},
\end{equation}
where the $\arctan$ terms are the local orientations of the external interface at points $J$ and $K$.

Let us denote $\sigma_{iT,iR}$(P) as the two-way travel time that considers refraction between layers.  Hence, we obtain the relationship:
\begin{equation}
\label{chap5:eq:delay_refr}
\begin{aligned}
    \sigma_{iT,iR}(P) = &\frac{\sqrt{(x_J - x_t)^2+(z_J - z_t)^2}+\sqrt{(x_K - x_r)^2+(z_K - z_r)^2}}{V_1}+\\
    &\frac{\sqrt{(x_P - x_J)^2+(z_P - z_J)^2}+\sqrt{(x_P - x_K)^2+(z_P - z_K)^2}}{V_2}.%
\end{aligned}
\end{equation}
Point $J$ depends on both emitting element $iT$ and focal point $P$, point $K$ depends on both receiving element $iR$ and focal point $P$. After angular transformation using equation~(\ref{chap5:eq:angle_ref}), $J$ and $K$ can be obtained by finding the points that satisfy equations~(\ref{chap5:eq:refraction_transmit}) and (\ref{chap5:eq:refraction_receive}) respectively.  If the external interface is planar, analytic developments give solutions for point $J$ and $K$. However, if the external interface is not planar, solving these equations is not trivial. Solutions are found using numerical methods. We used MATLAB 2023a and its non-linear zero finding algorithm \textit{fzero} Copyright 1984-2021 The MathWorks, Inc.

Suppose that we have a curved reflector $(D_i)$ inside the cortex as in Figure~\ref{fig:refraction_reflection}. The reflector can be approximated by a parabola with parameters $b_0$, $b_1$ and $b_2$:
\begin{equation}
    \label{chap5:eq:internal_interface}
    (D_i): z=b_0+b_1x+b_2x^2
\end{equation}

The laws of specular reflection at this interface can be written as: 
\begin{equation}
    \label{chap5:eq:snell_descates_bone}
    \gamma_t +\gamma_r-2\theta_l=0,
\end{equation}
where $\theta_l$ is the local orientation of the interface passing through focal point $P$ and its value is given by $-\arctan(2b_2x_P+b_1)$. 

\subsection{Specular signature and specular transform in presence of refraction}
\begin{figure}[htb!]
                \begin{subfigure}{.3\textwidth}
                    \includegraphics[trim={0 102 0 102},clip,width=\textwidth]{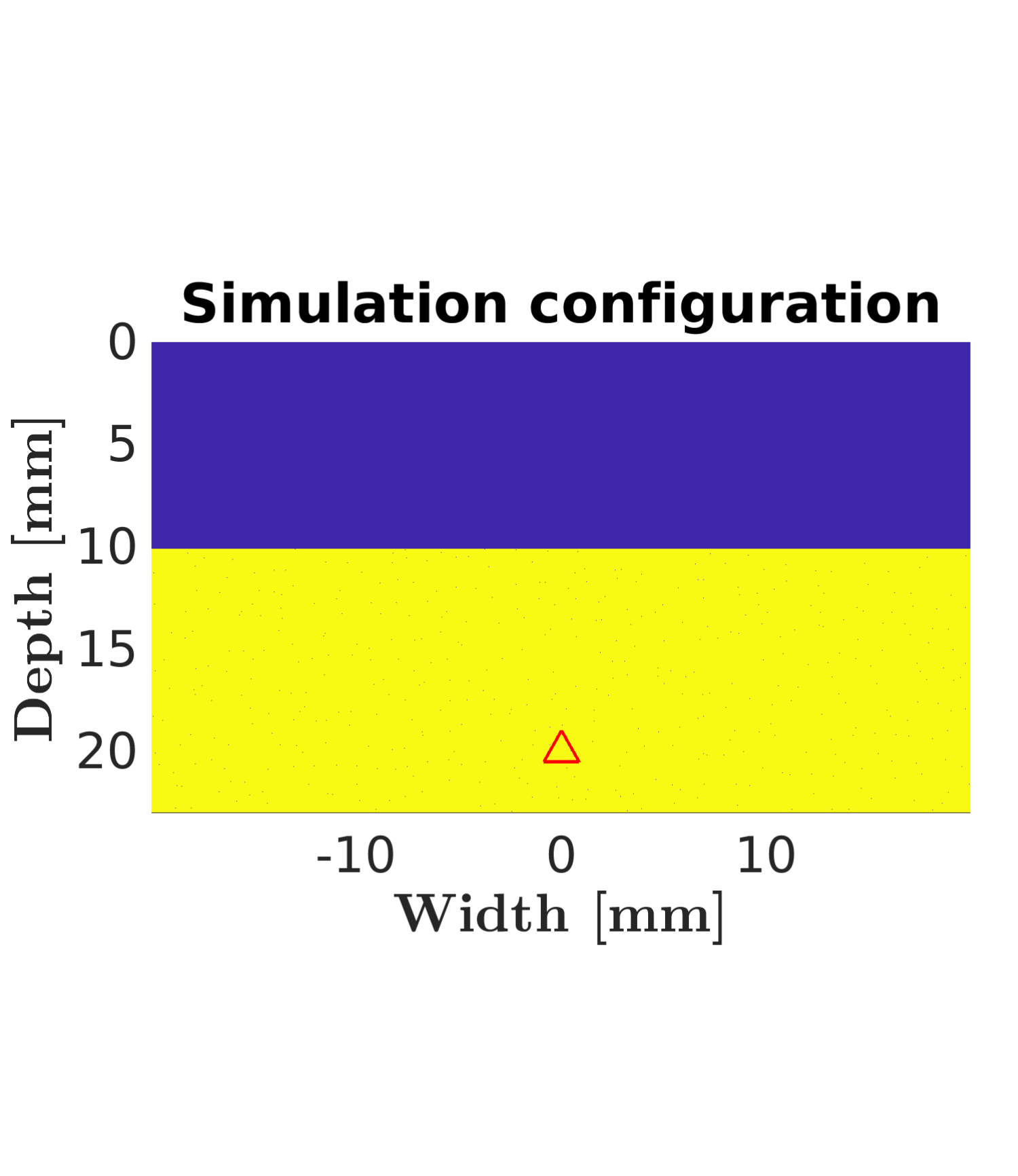}
                    \caption{}
                    \label{signature_bone:subfig:config_speckle}
                \end{subfigure}
                \begin{subfigure}{.3\textwidth}
                    \includegraphics[trim={0 0 0 0},clip,width=\textwidth]{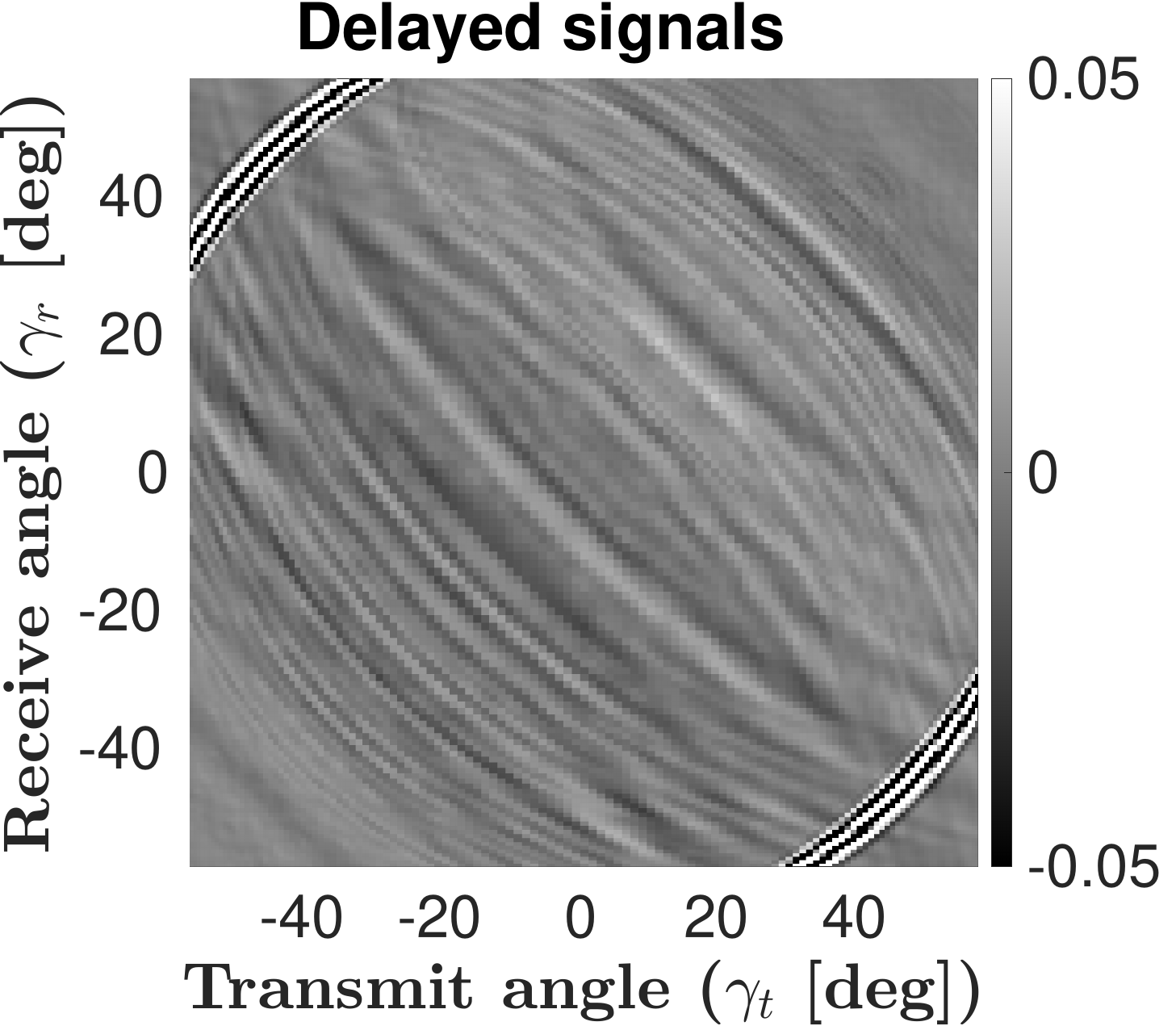}
                    \caption{}
                    \label{signature_bone:subfig:signature_speckle}
                \end{subfigure}
                \begin{subfigure}{.3\textwidth}
                    \includegraphics[trim={0 0 0 0},clip,width=\textwidth]{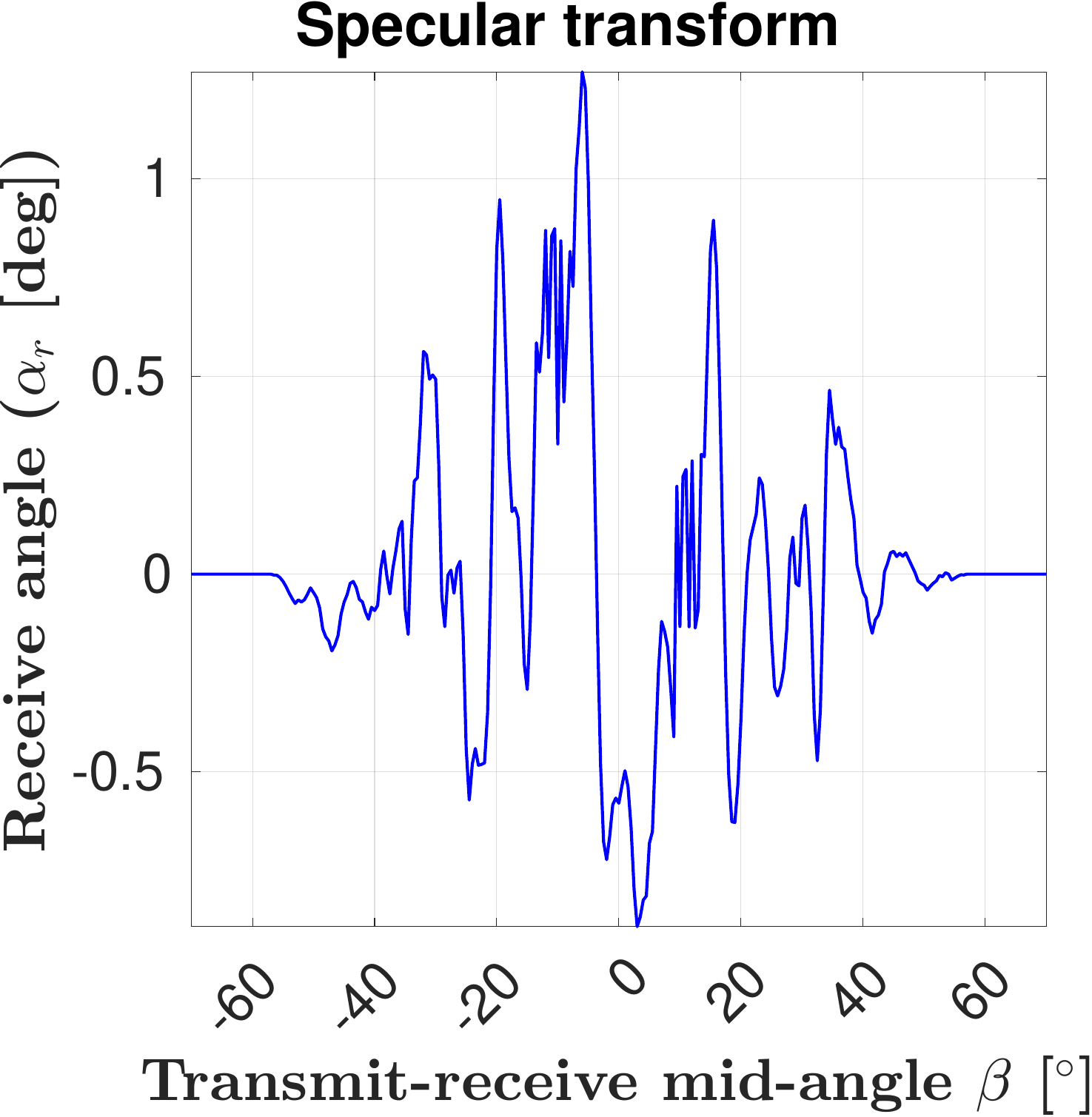}
                    \caption{}
                    \label{signature_bone:subfig:specutrans_speckle}
                \end{subfigure}
                \begin{subfigure}{.3\textwidth}
                    \includegraphics[trim={0 102 0 102},clip,width=\textwidth]{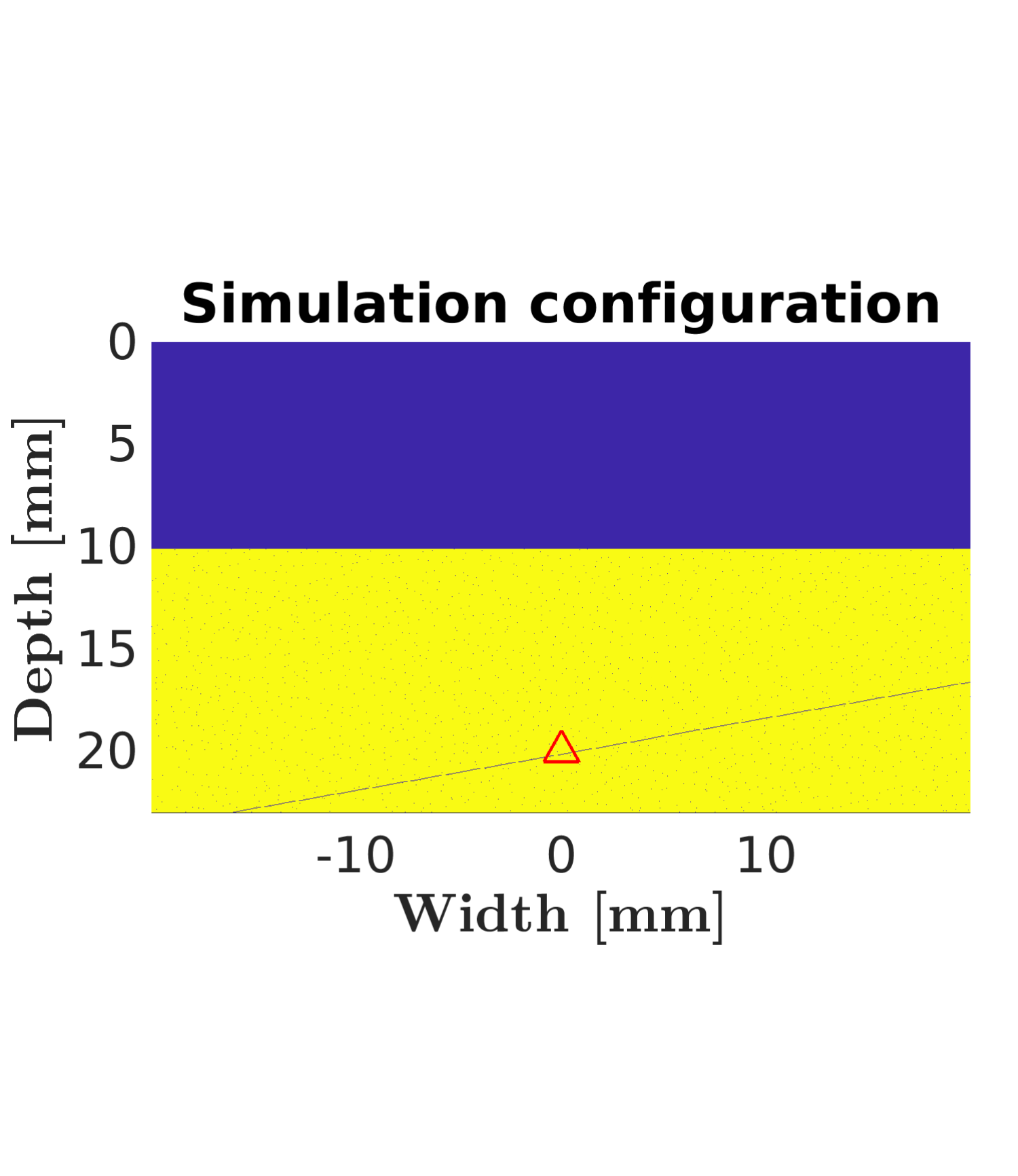}
                    \caption{}
                    \label{signature_bone:subfig:config_noisy_specular}
                \end{subfigure}    
                \begin{subfigure}{.3\textwidth}
                    \includegraphics[trim={0 0 0 0},clip,width=\textwidth]{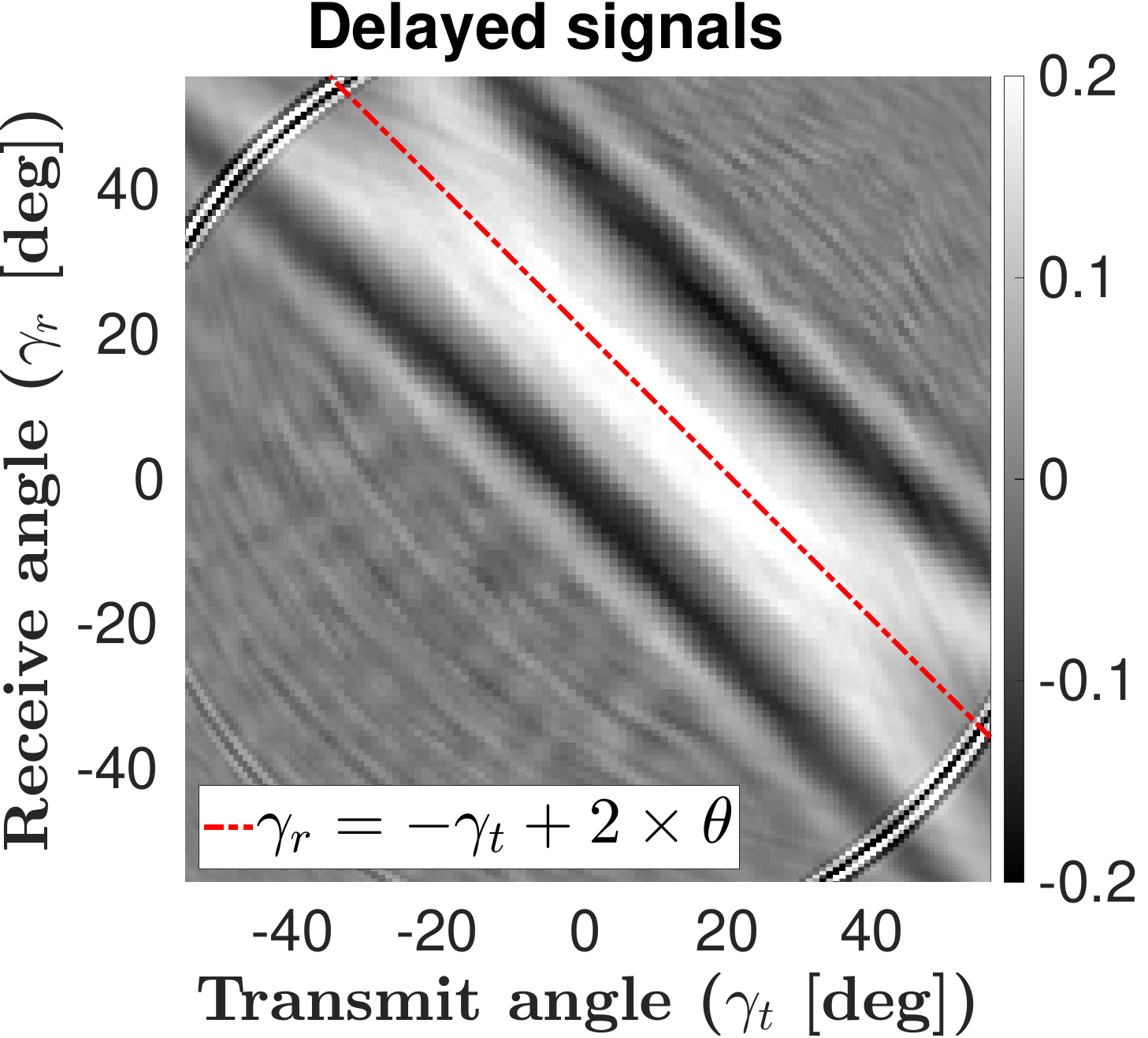}
                    \caption{}
                    \label{signature_bone:subfig:signature_noisy_specular}
                \end{subfigure}
                \begin{subfigure}{.3\textwidth}
                    \includegraphics[trim={0 0 0 0},clip,width=\textwidth]{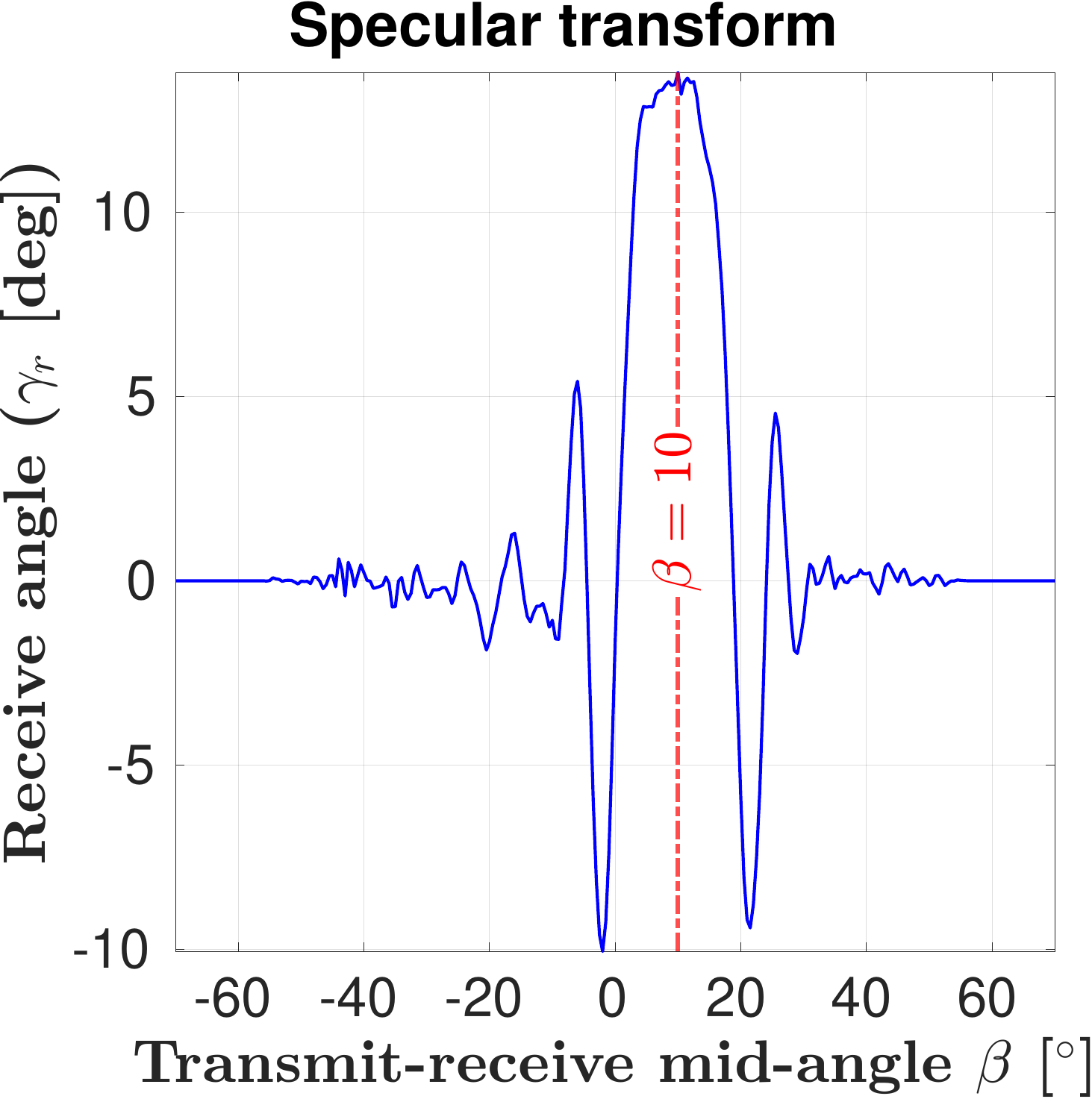}
                    \caption{}
                    \label{signature_bone:subfig:specutrans_noisy_specular}
                \end{subfigure}
                \caption{Delayed received signals from a pixel located at depth (0,20)mm inside the second layer (red triangle in left column images) for all different configurations. The first layer is water with speed of sound of 1540 m/s and the second layer is bone matrix with speed of sound 3500 m/s. The left images (panels (a) and (d)) are the simulation configuration, the middle column images (panels (b) and (e)) are the corresponding delayed received signals with respect to receive and transmit angle at the pixel. The red line in panels (e) is the plot of the specular reflection equation~\ref{chap5:eq:snell_descates_bone} for $\theta$ = 10$^{\circ}$, and the right column images (panels (c) and (f)) are the corresponding refraction corrected specular transform computed using equation~\ref{chap5:eq:specu_transform_bone}.}
            \label{fig:signature_bone}
            \end{figure}

    \paragraph{Specular signature}
        We consider a two-layer medium with the inclusion of different reflectors in the second layer (see Figure~\ref{fig:signature_bone} panel (a) to (d)). 
        
        We performed simulations with SimSonic software (\cite{bossy_simsonic_nodate}). A synthetic aperture imaging sequence with an array transducer (central frequency 2.5~MHz and a 3~dB bandwidth of 80\%) of 128 elements of size 245~$\upmu$m  and with a spatial period (pitch) of 300~$\upmu$m was simulated.  The first layer is a homogeneous fluid medium with water speed of sound $V_1$=1540 m/s and the second layer is a homogeneous elastic medium with longitudinal wave speed of sound $V_2=3500~m/s$ mimicking bone matrix. The simulated probe is immersed inside the first layer and the interface between two layers is at 10 mm depth from the probe surface. For this configuration 2 scenarios were considered similarly to \cite{rodriguez-molares_specular_2017}: 
        \begin{itemize}
            \item a scenario to illustrate pure speckle noise, numerous point scatterers are randomly placed in the second layer, (Panel~\ref{signature_bone:subfig:config_speckle})
            \item a scenario to illustrate specular scattering drown into speckle noise, a specular object with an orientation $\theta=10^{\circ}$ is placed at 20 mm depth and is surrounded by numerous point scatterers randomly distributed, (Panel~\ref{signature_bone:subfig:config_noisy_specular})
        \end{itemize} 
        
        Recorded signals from a focal point $P=(0,20)~$mm for each configuration are shown in Figure~\ref{fig:signature_bone} in panels (b) and (e).  The same signature are observed as for an homogeneous medium \cite{rodriguez-molares_specular_2017}.  

        In panels (b) and (e) of Figure~\ref{fig:signature_bone}, we observe in the upper left and lower right corners of the image the unwanted reflections from the external interface at 10$~$mm. These are the signals that follow an equivalent propagation path but do not originate from the wanted direction. These are contribution of secondary lobes and they can be eliminated with an appropriate f-number. 
        
        Apart from these artifacts and noise, we obtain the same specular patterns as in \cite{rodriguez-molares_specular_2017}. 
        \paragraph{Specular transform}
        From the above results, specular signature does not change when refraction is considered appropriately. We can therefore use the specular transform by considering the receive and transmit angle at the pixel ($\gamma_t,\gamma_r$). Hence, the same transformation as in \cite{rodriguez-molares_specular_2017}  translates the received signals in the specular domain through :
        \begin{equation}
        \label{chap5:eq:specu_transform_bone}
            f(\beta;P) = \sum_{iT=1}^{N_T}\textbf{S}(\sigma_{iT,iR}(P),\gamma_r,\gamma_t)|_{\frac{\gamma_r+\gamma_t}{2}=\beta}
        \end{equation}

        Panels (c) and (f) of Figure~\ref{fig:signature_bone} shows the specular transform of received signal for the configurations with pure speckle noise and for specular reflection with speckle noise.  After transformation, speckle noise is random (panel (c)) and specular reflection exhibits a certain shape specific to specular reflector (panel (f).

    \subsection{Model of specular transform}
        We derive a model of specular transform for a two-layered medium. We consider the reflector inside the second layer ($D_i$) as a parabola given by equation~(\ref{chap5:eq:internal_interface}) (Figure~\ref{fig:refraction_reflection}).
        The parameter $b_2$ determines the curvature of the reflector.

        When reaching the interface between the layers ($D_e$), part of the incident ray is reflected and another part is refracted inside bone cortex. At the reach of the specular reflector, the refracted ray is reflected and element $iR$ records the back scattered signal after another refraction. Therefore, there exists a unique point $Q\equiv(x_Q,z_Q)$ belonging to interface $D_i$ such that the transmit and receive angles at point $Q$ will satisfy the law of specular reflection and refraction given in equations~(\ref{chap5:eq:refraction_receive}),~(\ref{chap5:eq:refraction_transmit}) and ~(\ref{chap5:eq:snell_descates_bone}) (blue plain ray path in figure~\ref{fig:refraction_reflection}). This point noted $Q$ is the mirror point (Figure~\ref{fig:refraction_reflection}). 

        Hence, the specular travel time that considers refraction is given to delay equation~\ref{chap5:eq:delay_refr} applied to point $Q$: $\sigma_{iT,iR}(Q)$.  Recording of specular reflection will start at this time. The ray travel time is given by $\sigma_{iT,iR}(P)$.  The specular contribution at point $P$ is the shifted echo $e(\sigma_{iT,iR}(P)-\sigma_{iT,iR}(Q))$ and the model can thus be obtained by applying equation~\ref{chap5:eq:spec_model_bone_exact}.  Note that the travel time now also depends on the parameters of the external interface ($a_0,a_1,a_2$). 
        
        Applying the laws of specular reflection and refraction, we know that the coordinates of mirror point $Q$ depends on the parameters of the external interface $D_e$, on the coordinates of the transmitting and receiving elements and it also belongs to the reflector $D_i$. Mathematically, this means:
        \begin{equation}
            \label{chap5:eq:syst_mirror_point_bone}
            \left\{
            \begin{array}{ll}
                 z_Q= b_0+b_1x_Q+b_2x_Q^2 \\
                 \gamma_r(Q)= -2\arctan(2b_2x_Q+b_1)-\gamma_t(Q)\\
                  \frac{\sin(\alpha_t(P)+\arctan(2a_2x_J+a_1))}{V_1} = \frac{\sin(\gamma_t(P)+\arctan(2a_2x_J+a_1))}{V_2}\\
                  \frac{\sin(\alpha_r(P)+\arctan(2a_2x_K+a_1))}{V_1} = \frac{\sin(\gamma_r(P)+\arctan(2a_2x_K+a_1))}{V_2},\\
            \end{array}   
            \right.
    \end{equation}
    
        where $J$ and $K$ are the incidents point of the incident and reflected wave respectively. Using the trigonometric relationship in equation~\ref{chap5:eq:angle_ref}, we can replace angles and solve the system~\ref{chap5:eq:syst_mirror_point_bone}. For the case of a single homogeneous medium and a planar reflector, analytical development of the coordinates of a mirror point could be found. In this present case of a multi-layer medium with curved specular interfaces, analytical developments are difficult to obtain. Hence, we use numerical computation to obtain $x_Q$ and $z_Q$.  We solved the non-linear system~\ref{chap5:eq:syst_mirror_point_bone} using  MATLAB 2023a and its non-linear zero finding algorithm \textit{fzero} (Copyright 1984-2021 The MathWorks, Inc) \textbf{THE CODE IS PROVIDED AS SUPPLEMENTARY MATERIAL}.
        
\begin{figure}[htb!]
            \begin{subfigure}{.495\linewidth}
                \includegraphics[width=\linewidth]{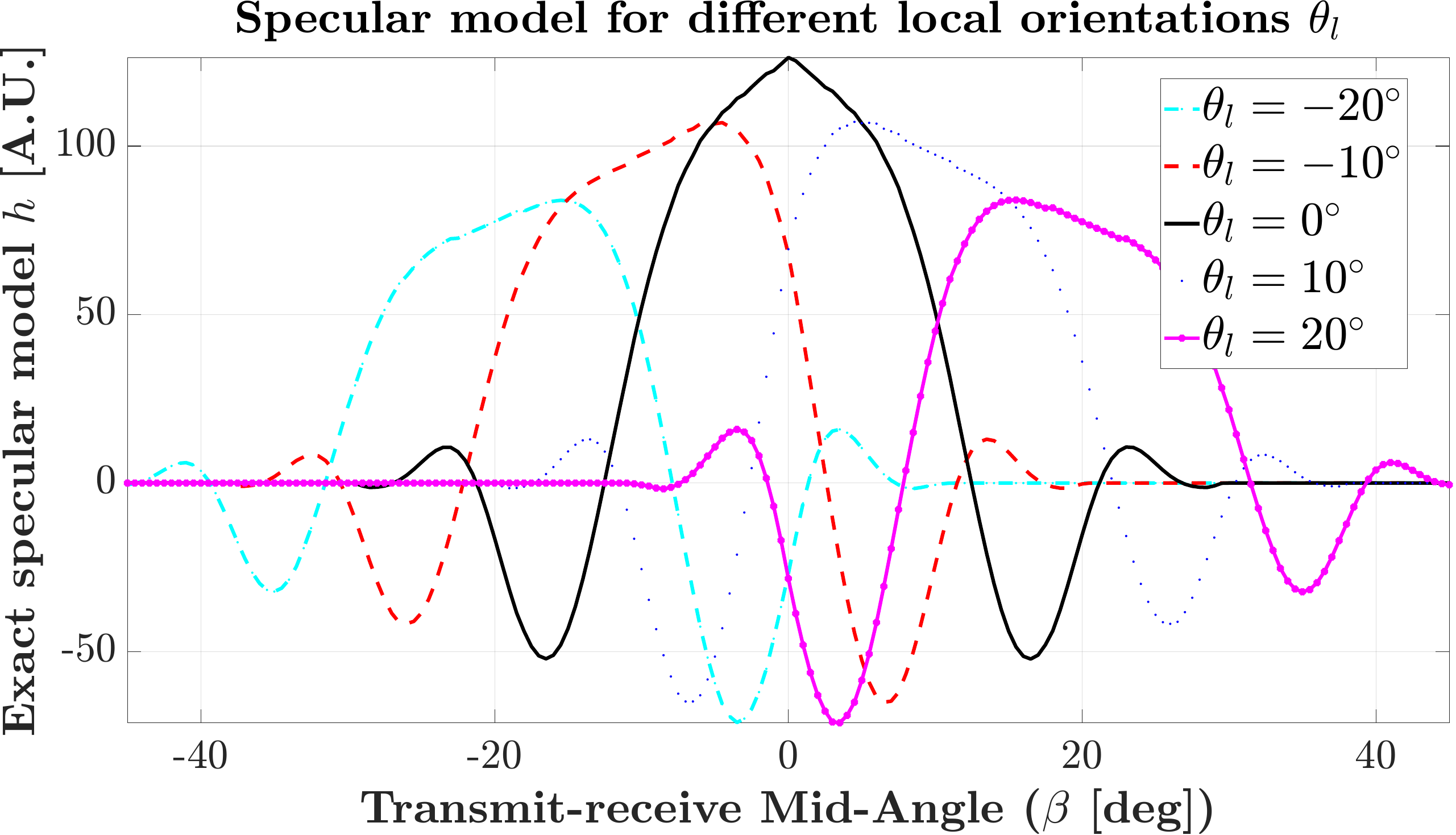}
                \caption{}
            \end{subfigure}
            \begin{subfigure}{.495\linewidth}
                \includegraphics[width=\linewidth]{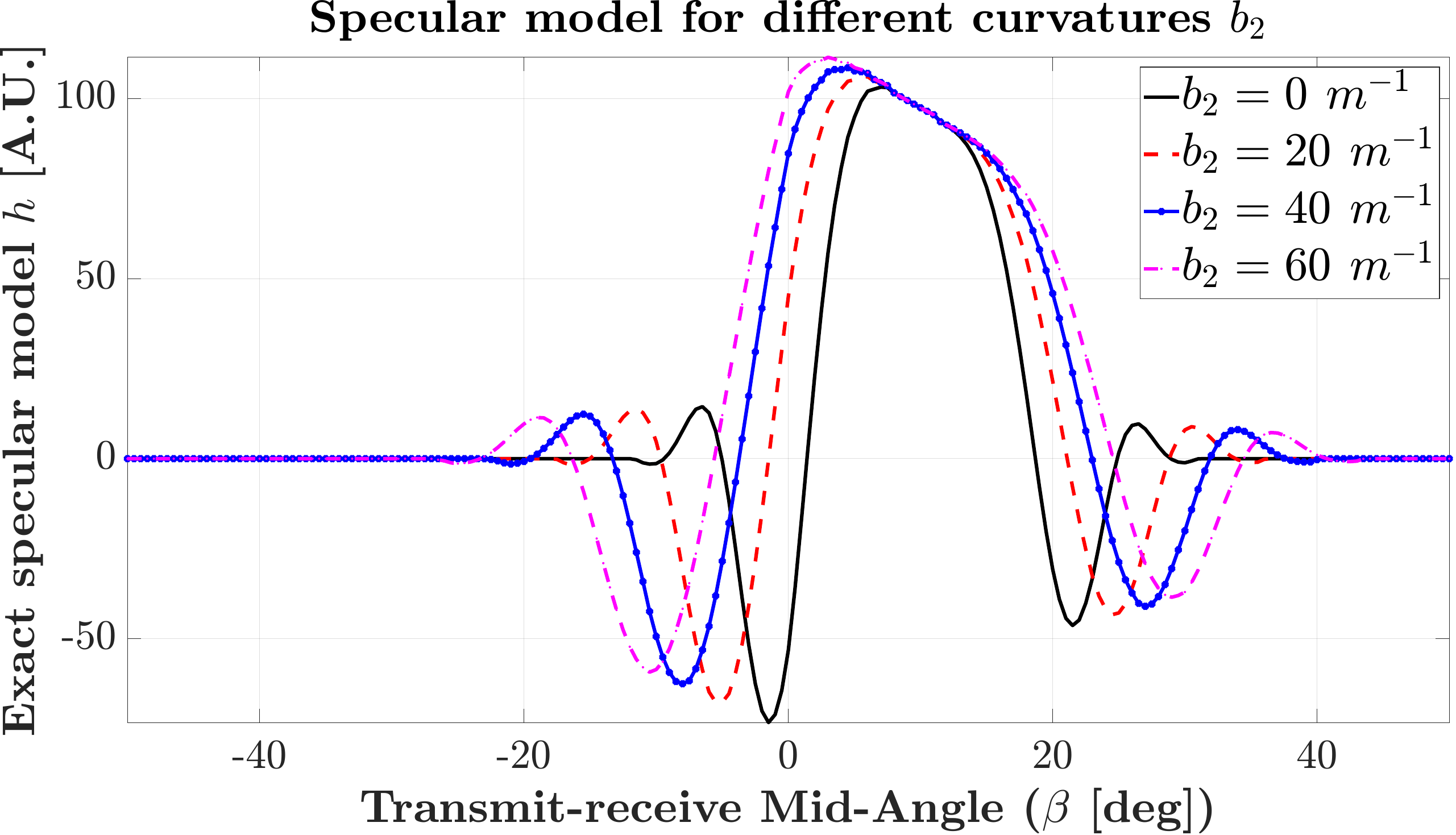}
                \caption{}
            \end{subfigure}
            \caption{Specular model for a specular interface inside the cortex with fixed curvature ($b_2=30~$mm$^{-1}$) and varying local orientations $\theta_l$ (panel (a)) and fixed local orientation ($\theta_l=10^{\circ}$ corresponding to $b_1=-0.17$) and varying curvatures $b_2$ (panel b). Parabolic parameters of the external interface are:  $a_0=10~$mm, $a_1=0$, $a_2=30 $mm$^{-1}$.}
            \label{chap5:fig20:model_bone_planar}
        \end{figure}

        In panel (a) of figure~\ref{chap5:fig20:model_bone_planar}, we plot the specular model obtained after computation of mirror points for 5 different local orientations ranging from -20 to +20$^{\circ}$ and fixed curvature parameter $b_2$ ($b_2=10~$m$^{-1}$). As expected, this model is similar to the model obtained for a planar interface by \cite{rodriguez-molares_specular_2017}. 

        In panel (b) of Figure~\ref{chap5:fig20:model_bone_planar} the local orientation of the specular interface is fixed to $10^{\circ}$, and specular models are plotted for 4 different curvatures of the reflector: $b_2=0~$m$^{-1}$, $b_2=20~$m$^{-1}$, $b_2=40~$m$^{-1}$ and $b_2=60~$m$^{-1}$. We observe that the spread of the specular model increases with curvature, but the center remains the same. Therefore, the curvature parameter $b_2$ alters the shape of the model. Similarly to the approach taken for needle tracking, we can derive a model of specular transform $h$ for a curved reflector with curvature $b_2$ and orientation $b_1$ passing through point $P$: 
        \begin{equation}
            \label{chap5:eq:spec_model_bone_exact}
            h(\beta;P,b_1,b_2) = \sum_{iT=1}^{N_T}\sum_{iR=1}^{N_R}e(\sigma_{iT,iR}(P)-\sigma_{iT,iR}(Q)).
        \end{equation}

        The parameter $b_1$ only shifts the simplified model but does not change its shape and $b_2$ changes the shape of the model. Hence, $h_0$ is invariant to $b_1$. We can then implement the matched filter by using the normalized cross-correlation between the model $h(\beta;P,b_1=0,b_2)$ and $f(\beta;P)$. 
        \begin{equation}
            \label{chap5:eq:cross_corr_bone}
            \chi(\theta_l;P,b_2) = \frac{\int_\beta{f(\beta;P)}\cdot h_0(\beta;P,0,b_2)d\beta}{\sqrt{\int_\beta{f(\beta;P)^2d\beta\cdot\int_\beta{h_0(\beta;P,0,b_2)^2}d\beta}}}.
        \end{equation}
We deduce the specularity index by taking the maximum correlation :
\begin{equation}
        \label{chap5:eq:specularity_bone}
        \Psi(P) = \max(\lVert\chi(\theta_l;P,b_2)\rVert).
    \end{equation}
    This maximum corresponds to a local orientation $\Tilde{\theta_l}$ and a curvature $\Tilde{b_2}$ given by : 
\begin{equation}
        \label{chap5:eq:orientation_bone}
        [\Tilde{b_2}, \Tilde{\Theta_l}](P) =\underset{b_2, \theta_l}{\arg\max}(\lVert\chi(\theta_l;P,b_2)\rVert).
    \end{equation}
    Estimate of the orientation parameter $\Tilde{b_1}$ can be deduced from the relationship $\tan\Tilde{\theta_l}=-(2\Tilde{b_2} x_P + \Tilde{b_1})$.

Similarly, one can generalize this procedure for a medium where the number of layers is above 2 and for other interfaces governed by higher degree polynomial. We then can get for any pixel, specular transform and a specular model that considers refraction. This allows to proceed to the model-based characterization of all reflections of the medium.

\section{Materials and methods}
\newcommand{\proxithird}{proximal third}
\subsection{Data acquisitions}
\subsubsection{\Exvivo~data}
    We used \exvivo~data described in \cite{dia_ultrasound_2025}. 
    Briefly, cortical bone samples were extracted from five human femoral shafts selected from a dataset comprising ten femurs obtained from female subjects aged between 66 and 98 years. The samples were immersed in water and scanned using a fully programmable ultrasound system (Vantage, Verasonics Inc., Redmond, WA, USA). The scanning scheme employed a synthetic aperture protocol (\cite{karaman_synthetic_1995, jensen_synthetic_2006}) in which each element in the array was sequentially activated, followed by a full array recording of the received echo signals. A phased-array ultrasound transducer with 96 elements operating at the central frequency of 2.5~MHz (P4-1 ATL/Philips, Bothell, WA, USA; pitch 0.295~mm) was used. The emitted pulse had a -3dB bandwidth of 1.33 MHz. A complete ultrasound acquisition resulted in a total of 96$\times$96 pulse-echo signals.  

    The samples are approximately 7 cm long, while the elevation of the probe is around 1.5~cm. Therefore, the sample is divided into 4 equal subvolumes along bone axis. Ultrasound recordings were obtained for a subvolume of every sample, with the setup illustrated in \cite{dia_ultrasound_2025}. The probe was positioned in front of the sample, slightly submerged in water. Acquisitions were repeated 10 times with repositioning, guided by real-time visualization to adjust the images accurately. Sample 4 from was excluded from this study due to its high heterogeneity, posing challenges in estimating ultrasonic wave speed. 

\subsubsection{\Invivo~data}
    This study was part of a study involving healthy subjects aimed at measuring intra-osseous blood flow in the tibia using ultrasound similar to \cite{salles_revealing_2021}. The research included eleven healthy male participants aged between 24 and 31 years old, with an average age of 28. Prior to participation, informed written consent was obtained from each participant for research purposes, in compliance with the legal requirements outlined in the French Code of Public Health approved by Ethics Committee and French Health Authorities (NO IDRCB: 2019-A02589-48, ClinicalTrials ID: NCT04396288).

    Ultrasound acquisitions were performed perpendicular to the bone axis in transverse planes. The measurement sites were located at the middle of the tibia (mid-diaphysis) and proximal one-third of the tibia (\proxithird). The length of the tibia was defined as the distance between the apex of the medial malleolus and the distal patellar apex. A pencil mark was made at each measurement site by the operators. Acquisitions were repeated five times at each site and for each measurement plane with repositioning, guided by real-time visualization. The probe alignment was confirmed when the ultrasound image presented bright external and internal interfaces.

    The same ultrasound system and phased array used in the \exvivo~study were employed for the \invivo~measurements. In this article, we choose the ultrasound image corresponding to the first ultrasound acquisition obtained at the proximal-third tibia. \textbf{Ultrasound image of the mid-diaphysis are provided as a supplementary material.}

    \begin{figure}[htb!]
        \centering
        \caption{Configuration for acquisition of ultrasound data \invivo. Probe is placed perpendicular to bone axis (transverse)}
        \label{fig:chap6:config_invivo}
    \end{figure}

\subsection{Ultrasound Image reconstruction}
\paragraph{DAS Beamforming}
DAS algorithm is used with a constant receive f-number of 0.5. This corresponds to a constant angular aperture at the transducer's elements of 45 degrees \cite{perrot_so_2021}. 
\paragraph{Specular Beamforming}
The specular algorithm describe in the teheory section produces three outputs: (1) a specularity map $\Psi$ that is the probability to find a specular structure at each pixel, (2) an orientation map $\Tilde{\Theta_l}$ that is an estimate of the most likely orientation of the specular structure and (3) an image that highlights specular structures and reduces speckle which we refer as specular beamformed image. Equation~\ref{cha5:eq:specular_beamforming} is used with a tolerance angle $\eta$ of 0.10 for planar interfaces and 0.25 for curved interfaces.

\begin{equation}
\label{cha5:eq:specular_beamforming}
I_{SP} (P) = \Psi(P)\cdot\sum_{\beta=\beta_{min}}^{\beta_{max}} w(\beta;\Tilde{\Theta_l}(P))\cdot f(\beta;P),
\end{equation}
where $w(\beta;\Tilde{\Theta_l}(P))$ denotes the apodisation window used for pixel $P$ (a hann window centered around $\Tilde{\Theta_l}(P)$):
\begin{equation}
\label{cha5:eq:specular_hann_window}
w(\beta;\Tilde{\Theta_l}(P)) = \left\{
\begin{array}{ll}
     &  \cos^2(\beta-\Tilde{\Theta_l}(P)), \ \ \ \ if\ |\beta-\Tilde{\Theta_l}(P)|\leq \eta\frac{\pi}{2}\\
     & 0 , \ \ \ \ if\ |\beta-\Tilde{\Theta_l}(P)|> \eta\frac{\pi}{2}
\end{array}
\right.
,
\end{equation}
here, $\eta$ is introduced as the specular tolerance, ranging from 0 to 1. When $\eta=0$, only the specular reflection from the orientation $\Tilde{\Theta_l}$ is taken into account. As $\eta$ increases, more specular signals with decreasing weights are considered and summed.

\paragraph{Speed of sound estimation }
For \invivo~data, the wave speeds in cutaneous tissues and in cortical bone estimated using an autofocus approach described in \cite{renaud_vivo_2018} are used. For \exvivo~data, the wave speed in cutaneous tissues was estimated using the head wave propagating at the interface between the probe and water. The wave speed in cortical bone was estimated using autofocus, as described and reported in \cite{dia_ultrasound_2025} .

\subsection{Endosteal interface visibility quantification}
\begin{figure}[htb!]

\includegraphics[width=.45\textwidth]{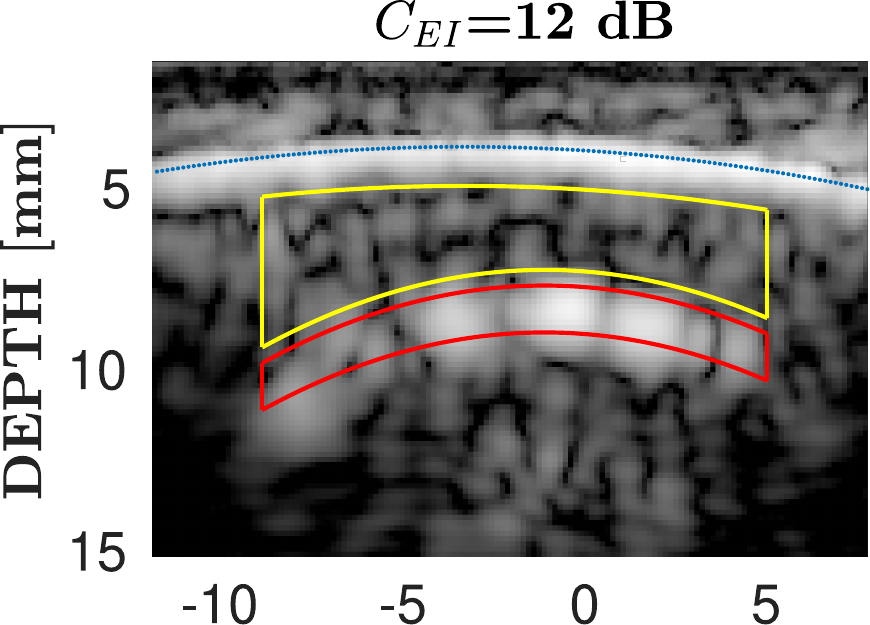}
\caption{Region of interest to quantify endosteal interface visibility. Reconstructed image from \invivo~data. The yellow and red ROIs are used to evaluate inner bone cortex and endosteum contrasts, respectively}
\label{chap6:roi_metrics}
\end{figure}

To evaluate the visibility of the endosteal surface, we use the endosteal interface contrast ($C_{EI}$) defined in \cite{dia_influence_2023} given by:
\begin{equation}
C_{EI} = \frac{\mu_E}{\mu_I},
\end{equation}
where $\mu_I$ and $\mu_E$ are the average image intensities in the center of the cortex and at the endosteal interface, respectively. The regions of interest (ROIs) used to quantify these contrasts are depicted in Figure~\ref{chap6:roi_metrics}.

It is important to note that the values of $C_{EI}$ for both specular beamforming (BF) and Delay-And-Sum (DAS) images are computed from the segmentation of the external and internal interfaces performed on the ultrasound image produced by the DAS algorithm.

\section{Results}
 \newcommand{\imageProxiThird}[2]{images/specular_only_1D_exact_BONEWAVESPEED_SA_Trans_SUJET#1_Step#2_frame_01_wind_length_045mod_comp.pdf}        
\newcommand{\imageMidDiaph}[2]{images/specular_only_1D_exact_BONEWAVESPEED_SA_Trans_SUJET#1_Step#2_frame_01_wind_length_045mod_comp.pdf}     
\newcommand{\stepPT}{18}
\newcommand{\imageSpeculaityProxiThird}[2]{images/specularity_1D_exact_BONEWAVESPEED_SA_Trans_SUJET#1_Step#2_frame_01_comp.pdf}
\newcommand{\imageOrientationProxiThird}[2]{images/orientation_1D_exact__BONEWAVESPEED_SA_Trans_SUJET#1_Step#2_frame_01_comp.pdf}

\subsection{Ex-vivo results}
\begin{figure}[htb!]
\setlength\tabcolsep{1pt}
\renewcommand{\arraystretch}{1} 
\begin{tabular}{m{.15\linewidth}m{.223\linewidth}m{.223\linewidth}m{.223\linewidth}m{.223\linewidth}}
    & \centerline{\bf Sample 1} &\centerline{\bf Sample 2}&\centerline{\bf Sample 3}&\centerline{\bf Sample 5}\\
     \centerline{Porosity} & \centerline{5.0 \%} &\centerline{5.3 \%}&\centerline{11.5 \%}&\centerline{16.6 \%}\\
     \centerline{X-ray}                           
    &\includepageasasubfigure{1}
    &\includepageasasubfigure{13}
    &\includepageasasubfigure{5}
    &\includepageasasubfigure{17}\\
\end{tabular}
\renewcommand{\arraystretch}{3.5} 
\begin{tabular}{m{.15\linewidth}m{.223\linewidth}m{.223\linewidth}m{.223\linewidth}m{.223\linewidth}}
    DAS BF                            
    &\multirow{2}{*}{
     \begin{subfigure}{\linewidth}
    \includegraphics[width=\linewidth,page=1]{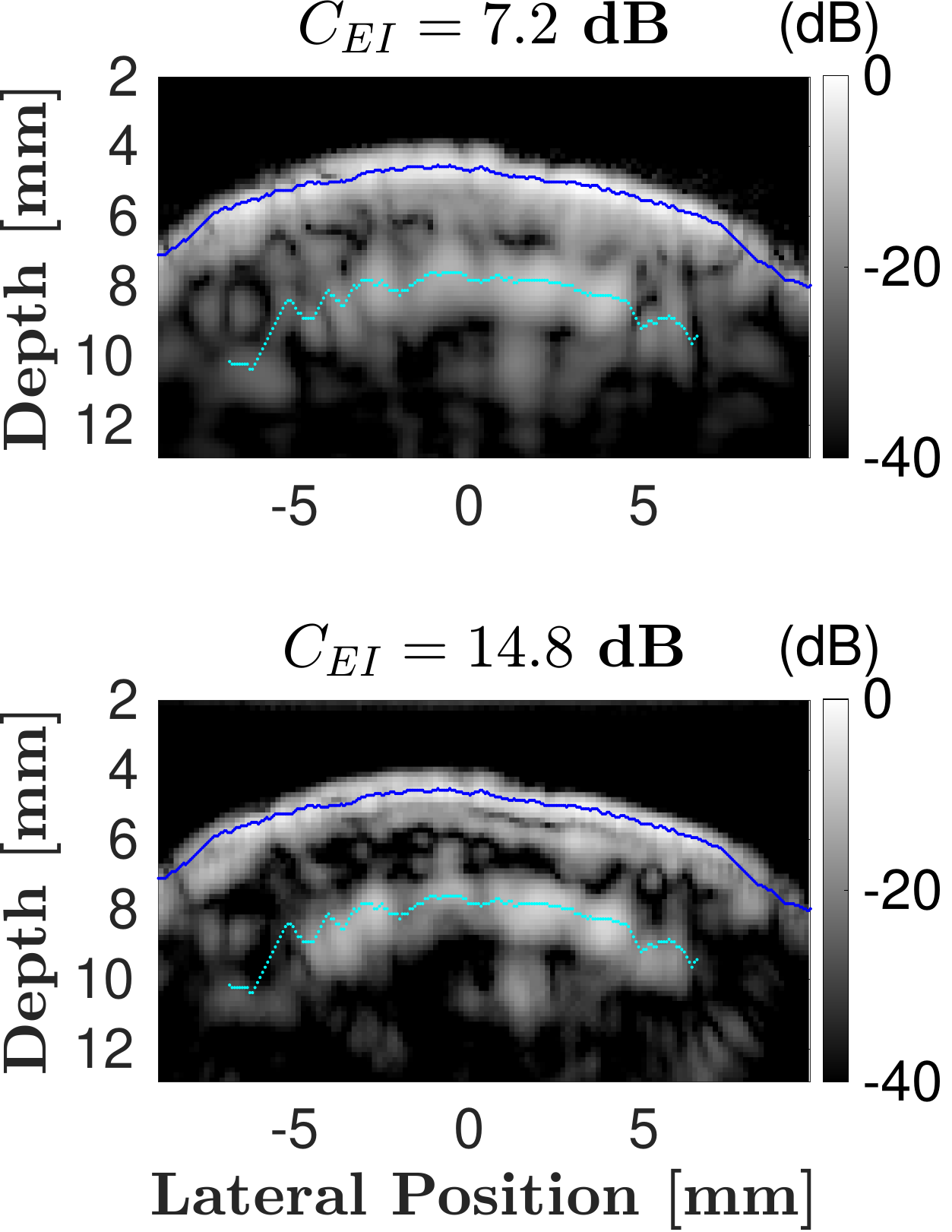}
     \end{subfigure}
    }
    &\multirow{2}{*}{
    \begin{subfigure}{\linewidth}\includegraphics[width=\linewidth,page=1]{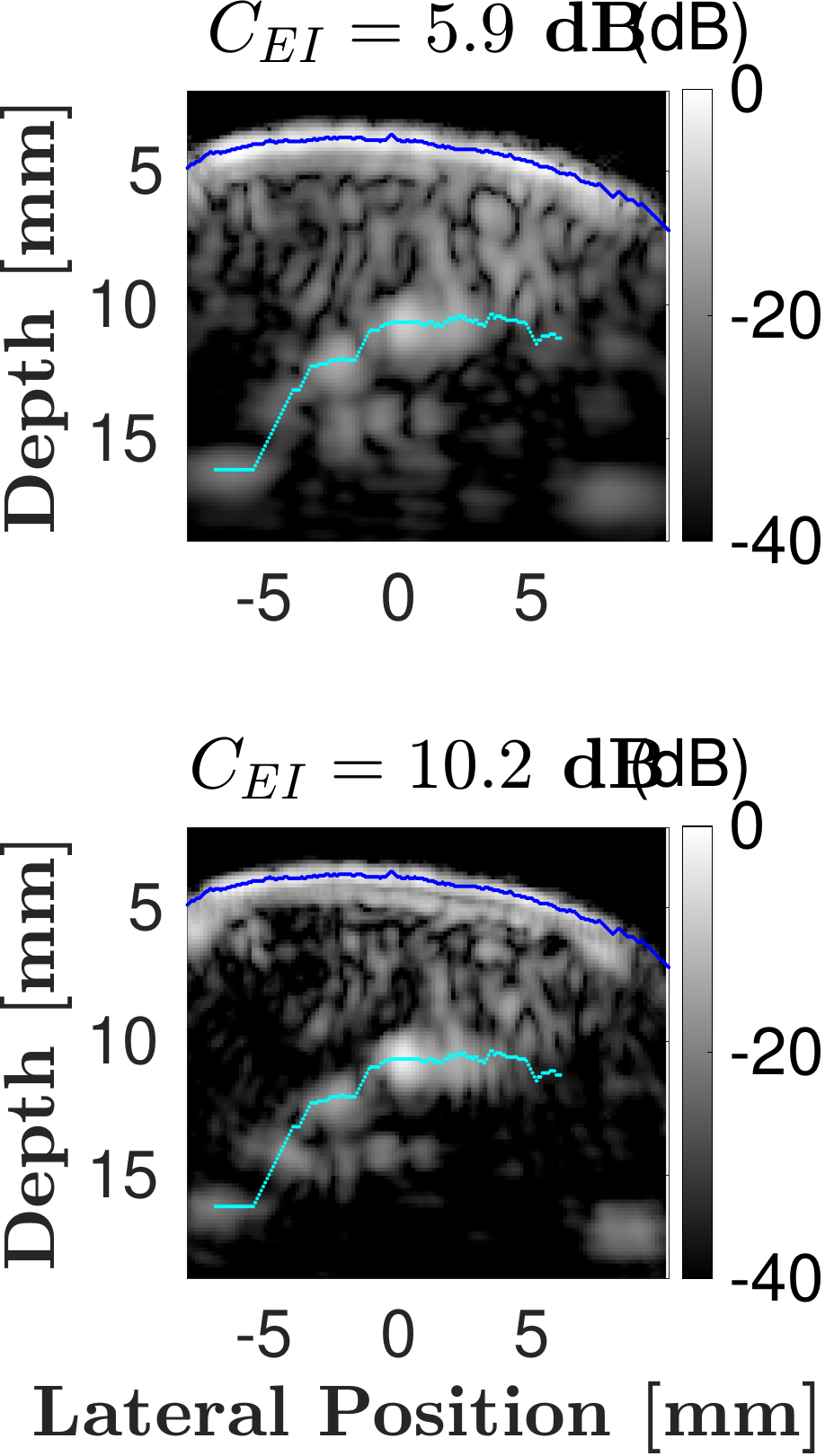}\end{subfigure}
    }
    &\multirow{2}{*}{
     \begin{subfigure}{\linewidth}
    \includegraphics[width=\linewidth]{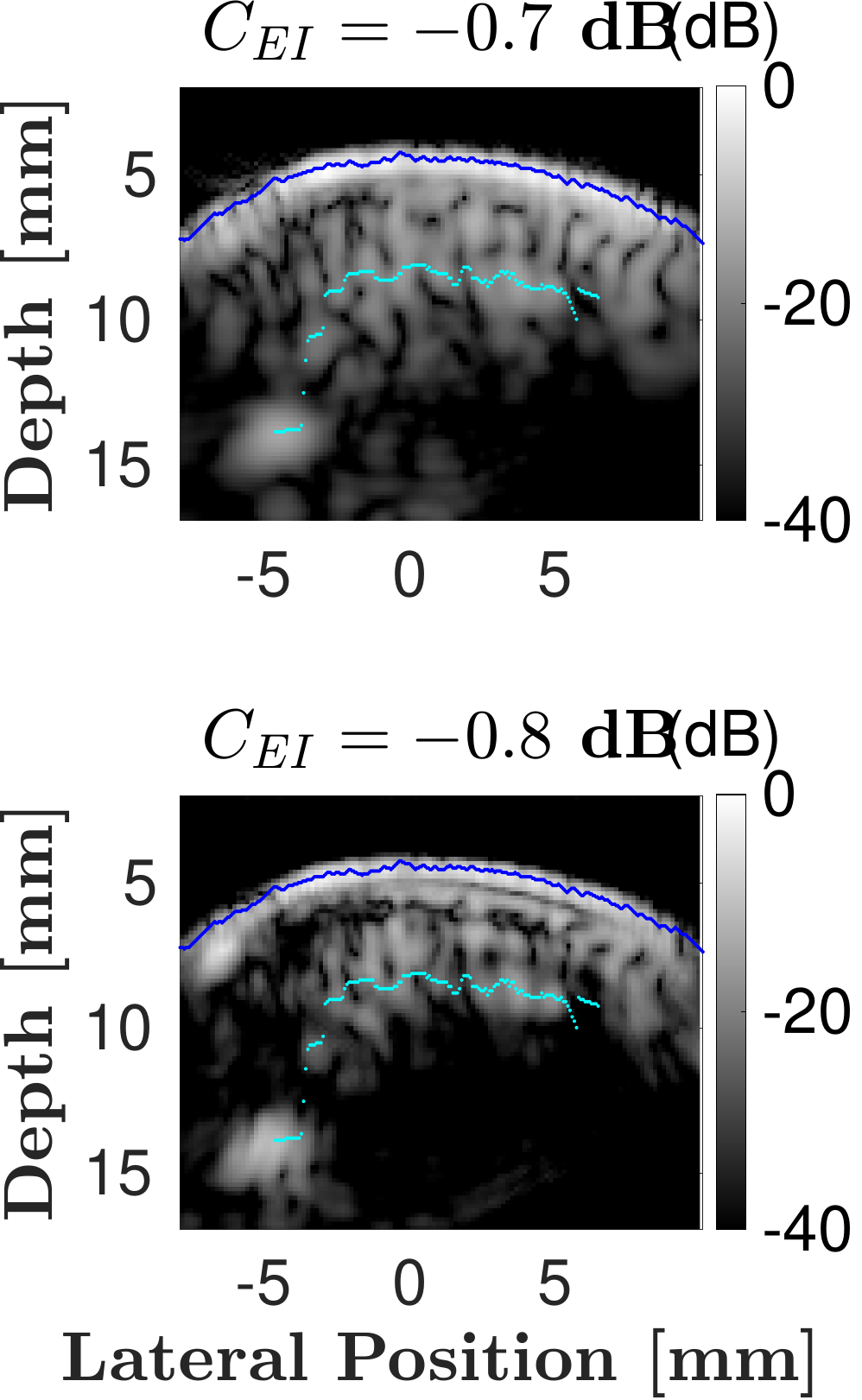}
     \end{subfigure}
    }
    &\multirow{2}{*}{
     \begin{subfigure}{\linewidth}
    \includegraphics[width=\linewidth,page=1]{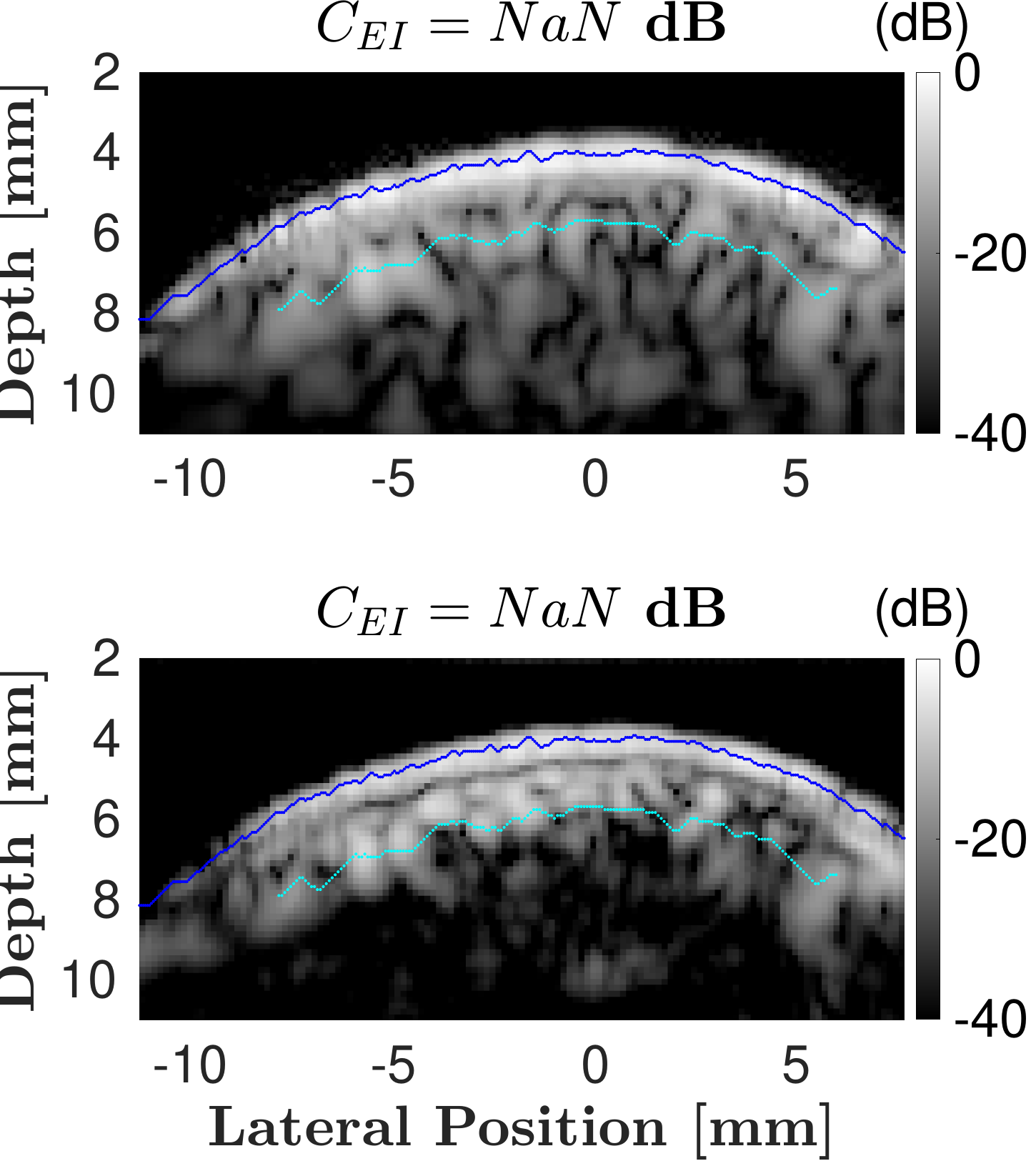}
     \end{subfigure}
    }
    \\
        Specular BF &
    &
    &
    &\\[2.5cm]
    \end{tabular}
\renewcommand{\arraystretch}{1} 
\begin{tabular}{m{.15\linewidth}m{.223\linewidth}m{.223\linewidth}m{.223\linewidth}m{.223\linewidth}}
    \centerline{Specularity} & 
    \begin{subfigure}{\linewidth}
    \includegraphics[width=\linewidth,page=1]{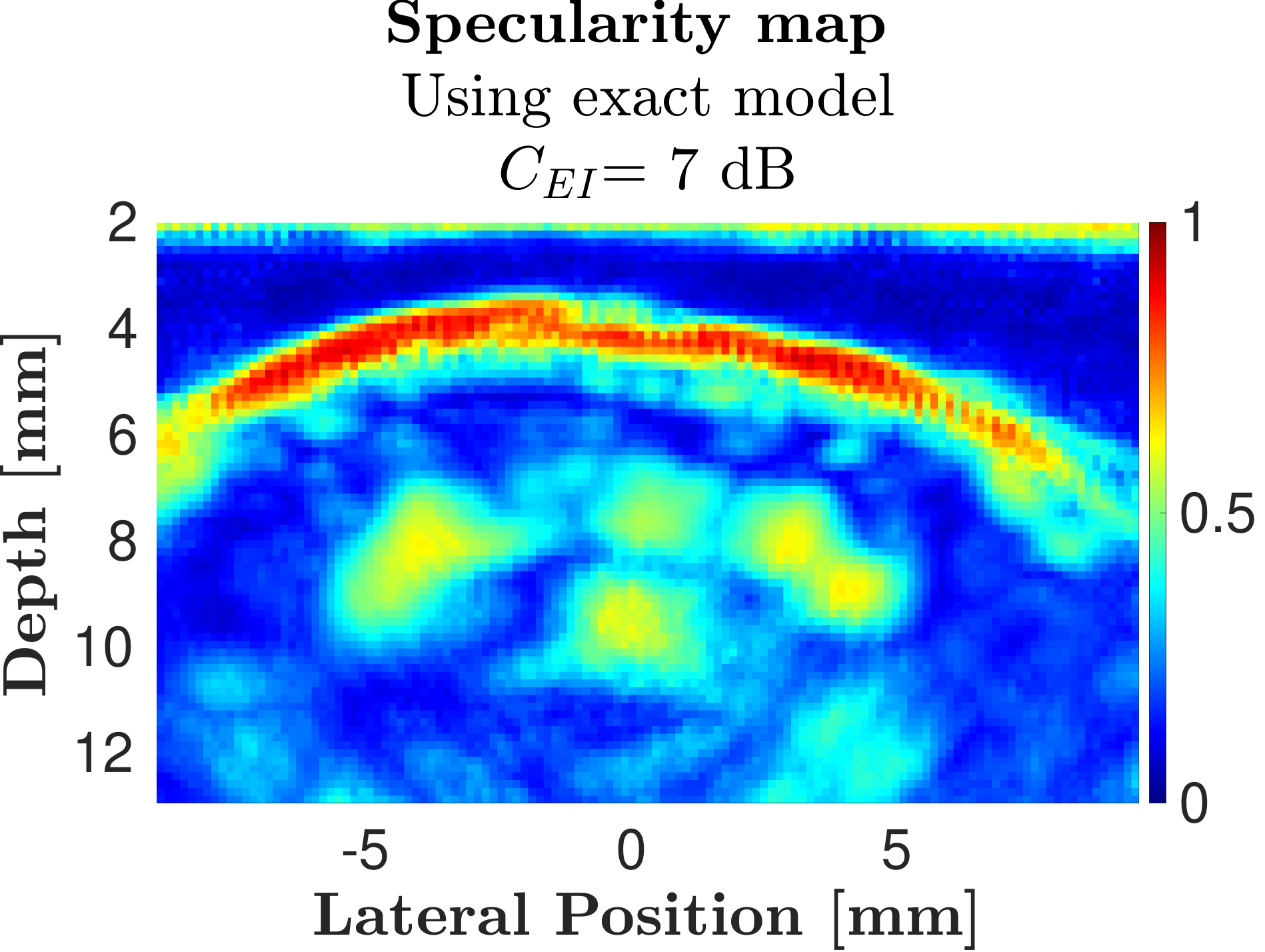}
    \end{subfigure}
    & 
    \begin{subfigure}{\linewidth}
    \includegraphics[width=\linewidth,page=1]{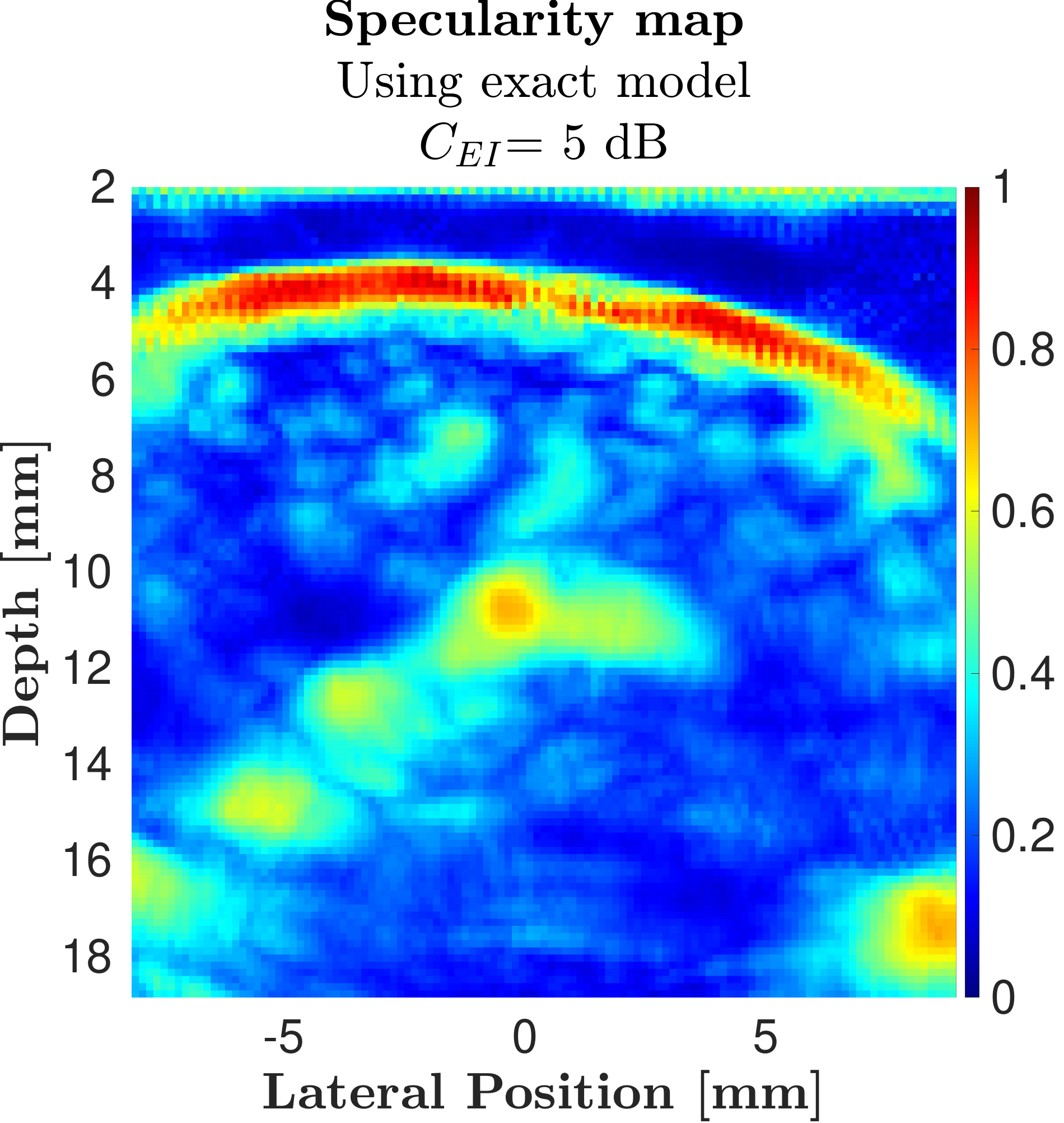}
    \end{subfigure}
    & \begin{subfigure}{\linewidth}
    \includegraphics[width=\linewidth,page=1]{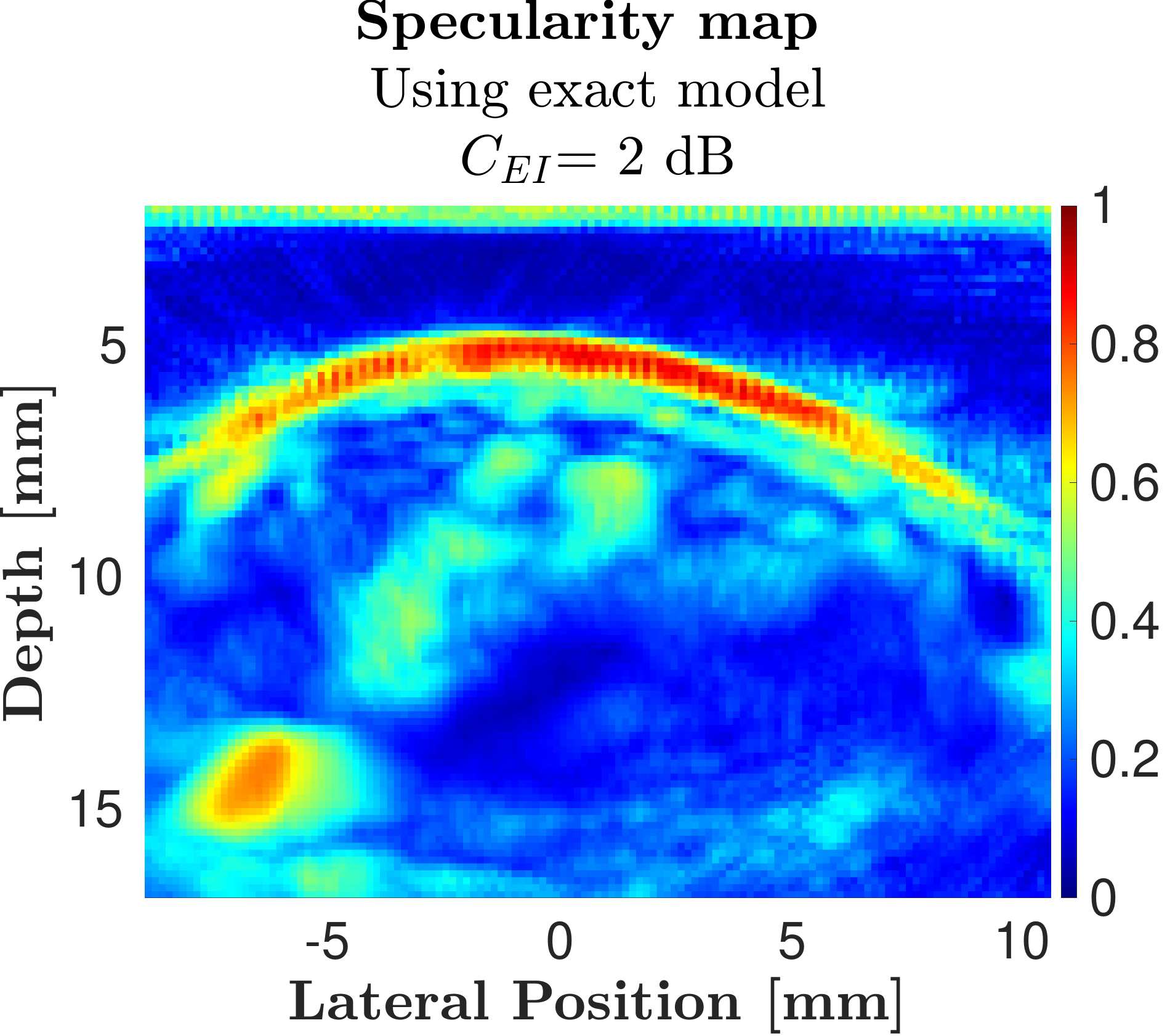}
     \end{subfigure}
    & 
     \begin{subfigure}{\linewidth}
    \includegraphics[width=\linewidth,page=1]{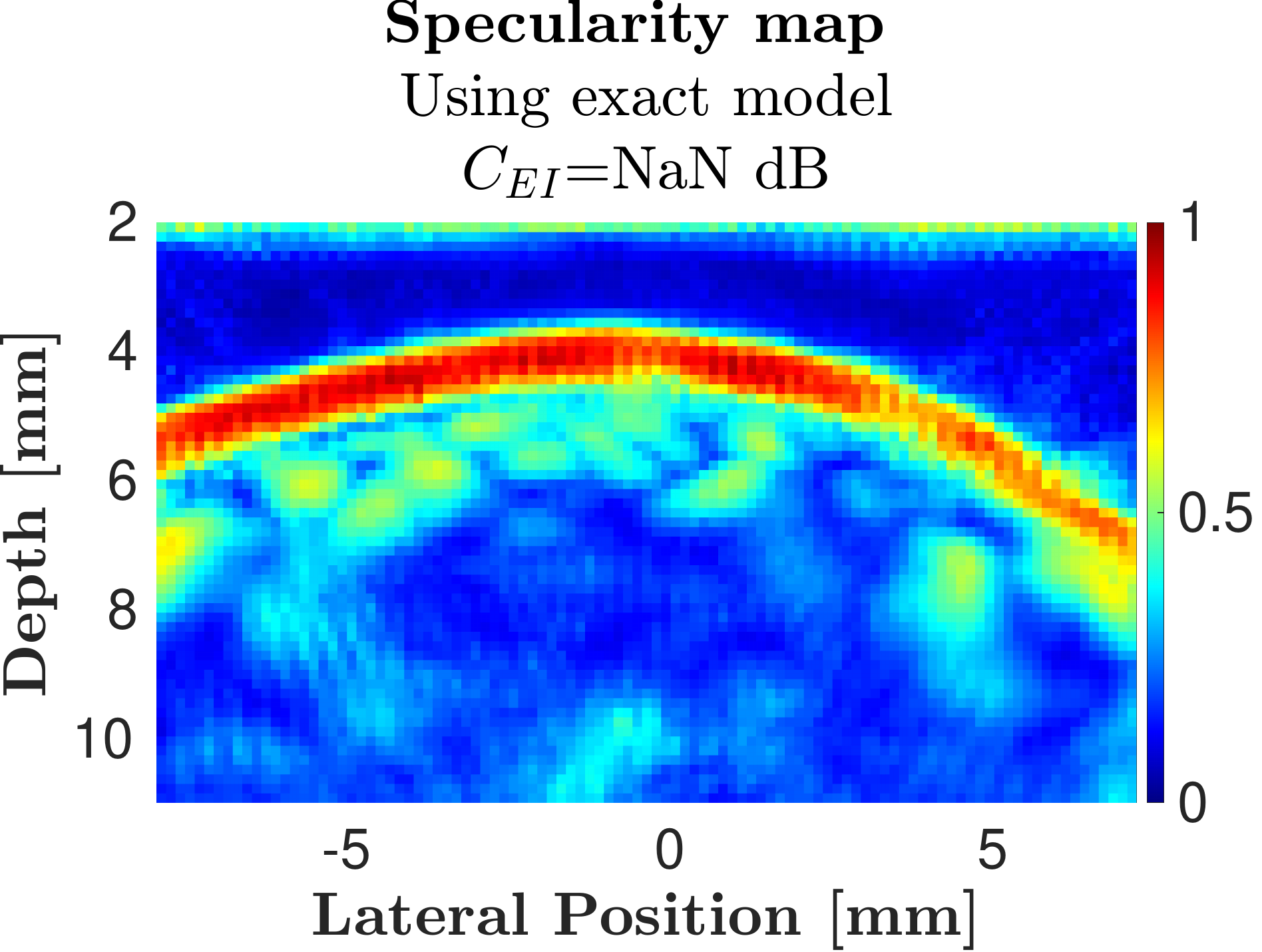}
     \end{subfigure}
    \\
    \centerline{Orientation}                            
    & \begin{subfigure}{\linewidth}
    \includegraphics[width=\linewidth,page=1]{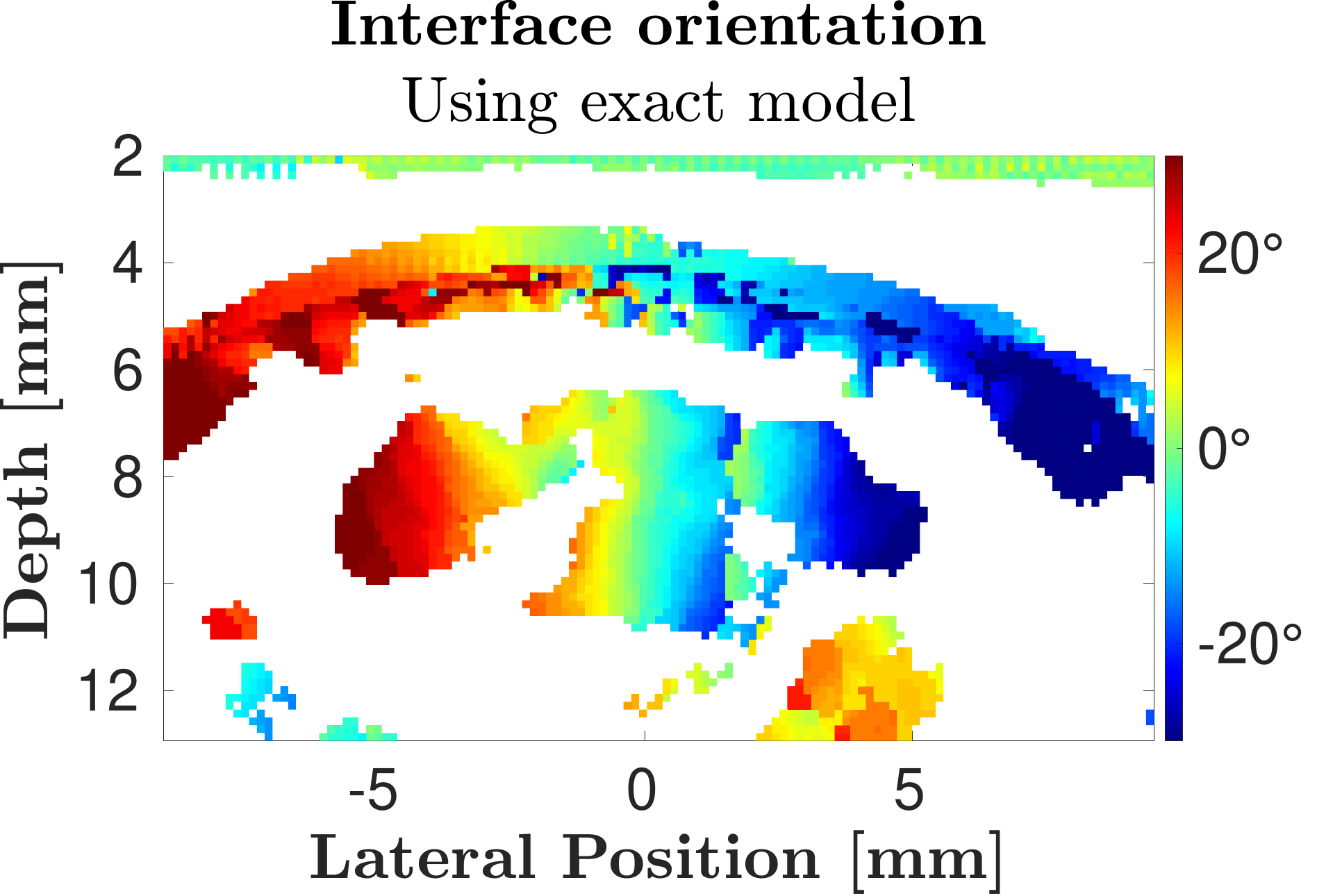}
     \end{subfigure}
    & \begin{subfigure}{\linewidth}
    \includegraphics[width=\linewidth,page=1]{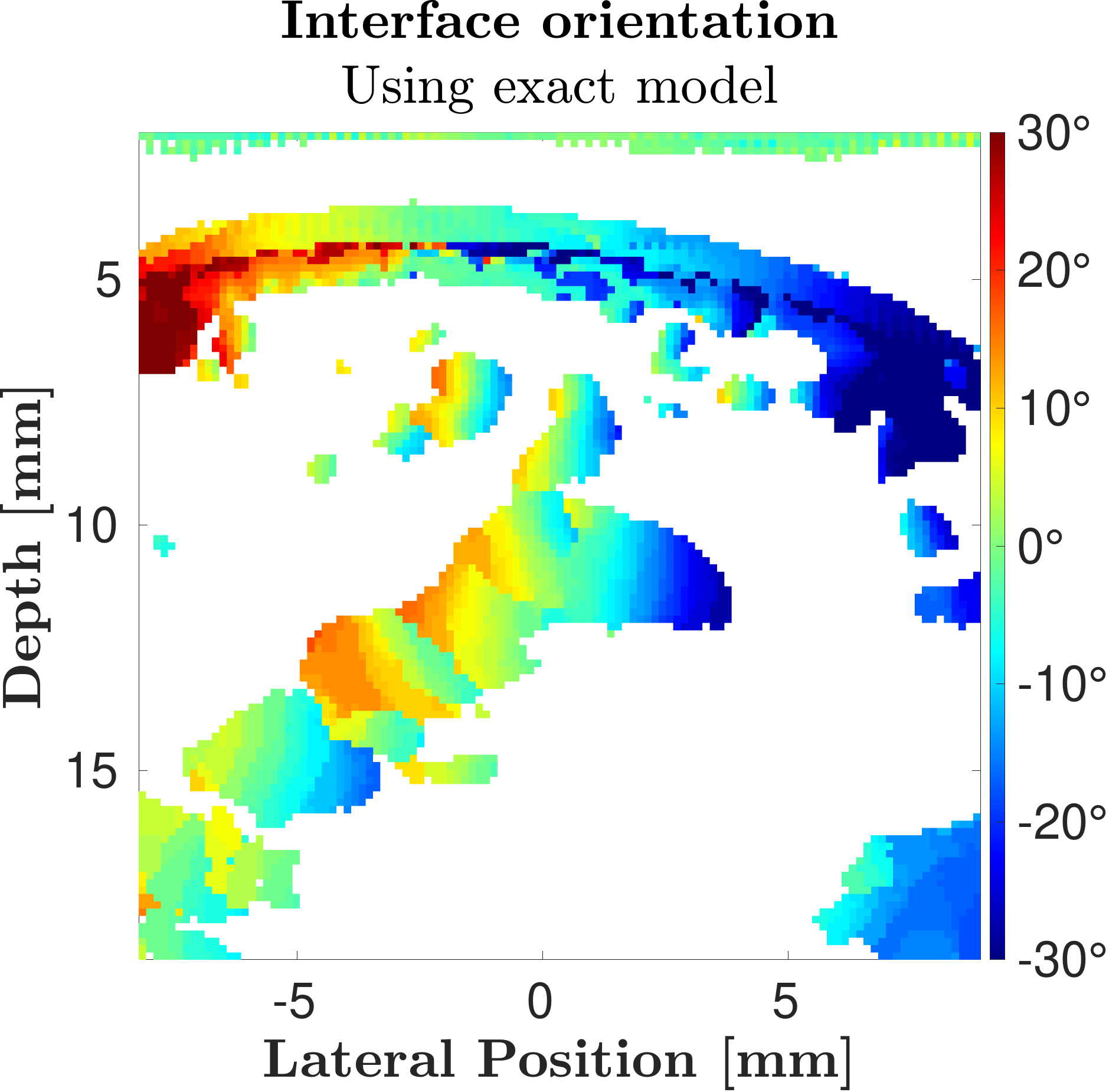}
     \end{subfigure}
    & \begin{subfigure}{\linewidth}
    \includegraphics[width=\linewidth,page=1]{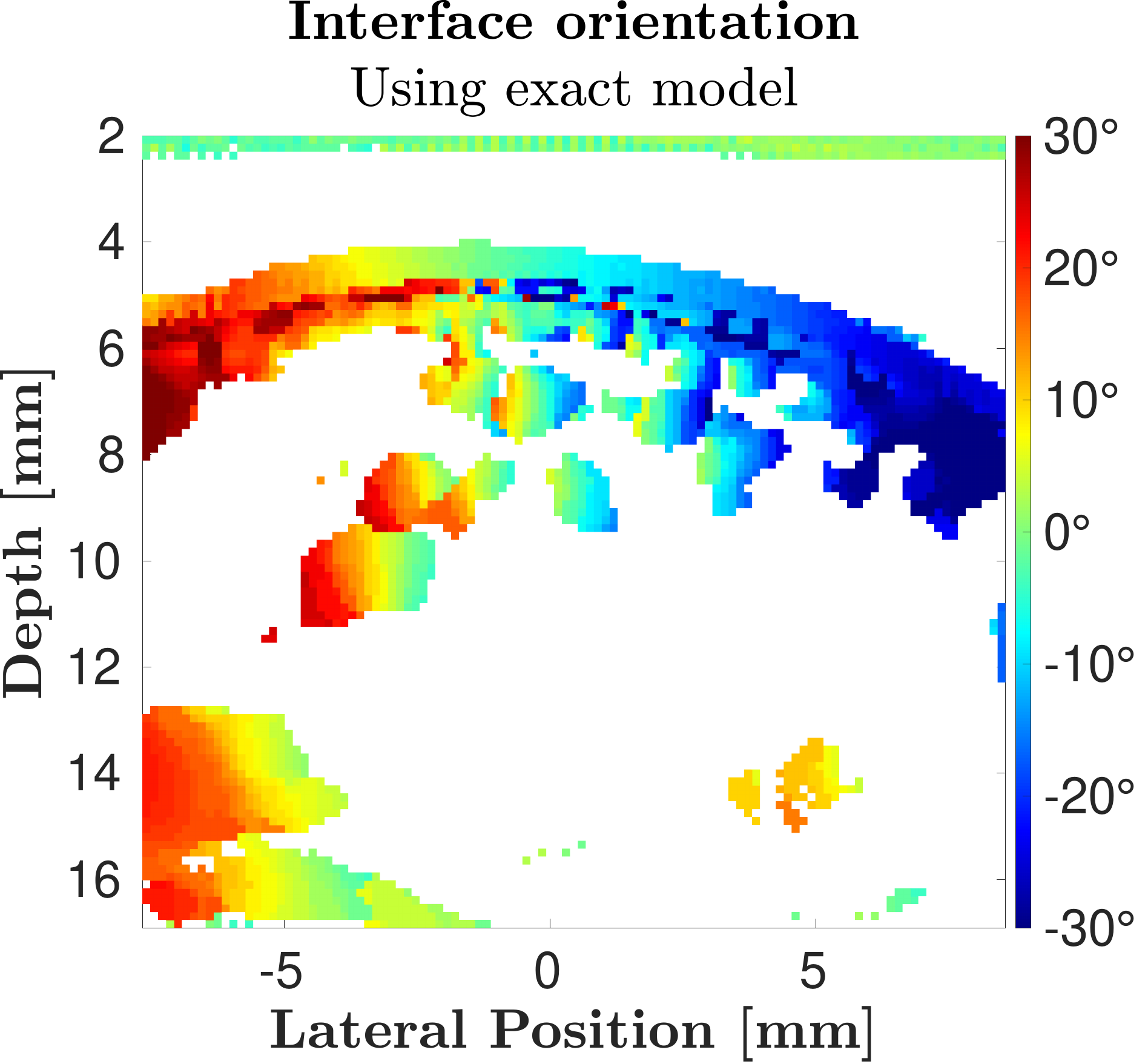}
     \end{subfigure}
    &  \begin{subfigure}{\linewidth}
    \includegraphics[width=\linewidth,page=1]{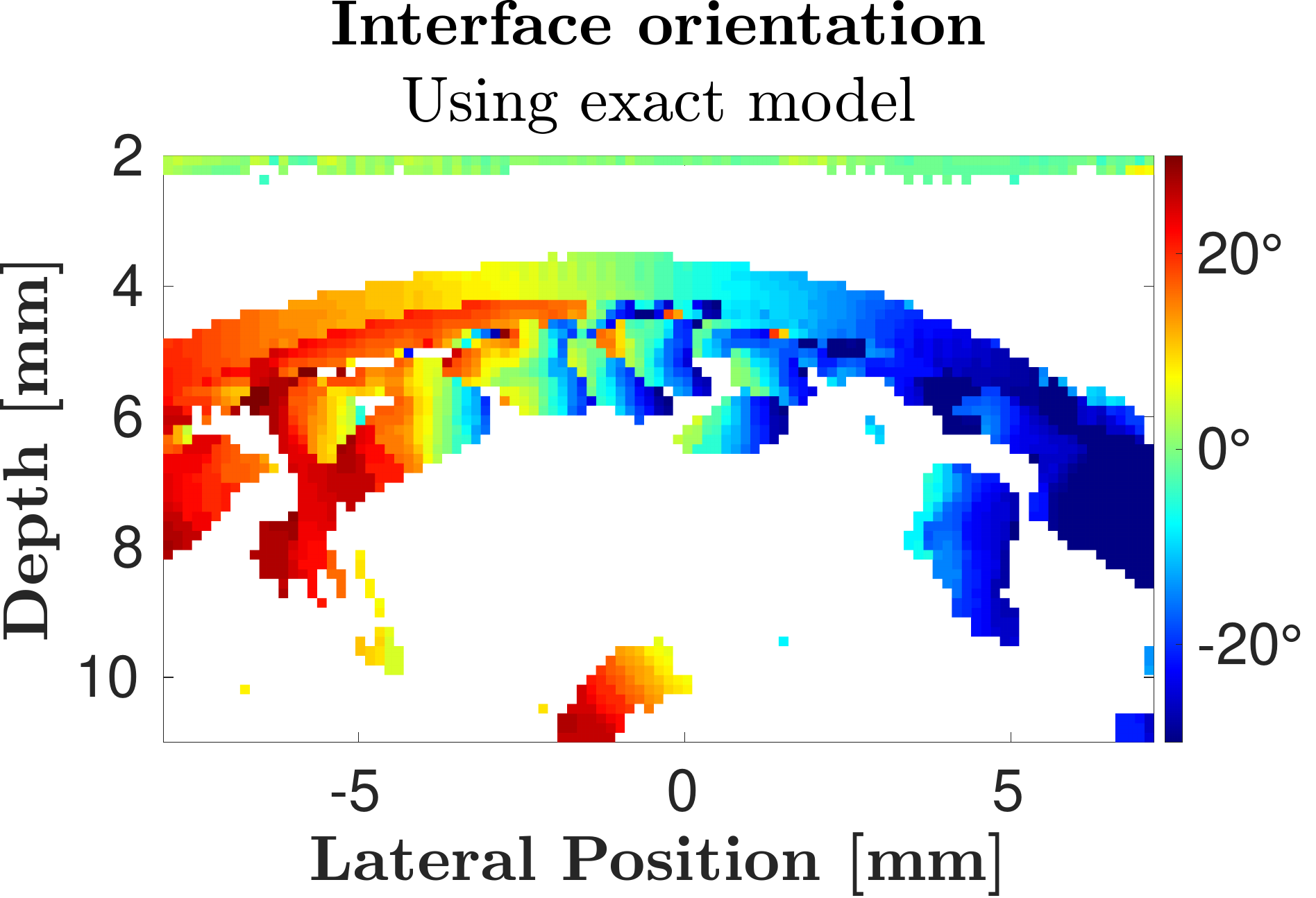}
     \end{subfigure}
    \\
    \end{tabular}
\caption{Comparison of ultrasound imaging techniques for each sample. Each column corresponds to subvolume 1 of each sample, and each row from top to bottom displays the 3D porosity of the subvolume, the X-ray micro-CT slice, the ultrasound image obtained with Delay-and-Sum beamforming (DAS BF), the ultrasound image obtained with Specular beamforming (Specular BF), the specularity map ($\Psi$), and the orientation map for pixels with specularity greater than 0.4}
    \label{fig_chap_6:results:sample_1}
\end{figure}

Figure~\ref{fig_chap_6:results:sample_1} presents the images obtained using Delay-and-Sum beamforming (DAS BF) and Specular beamforming (Specular BF) for subvolume 1 of samples 1, 2, 3, and 5. The corresponding images for all subvolumes of each sample are provided as supplementary material. 

Table~\ref{tab_chap6:summary_cei} summarizes the endosteal visibility metric ($C_{EI}$) computed for each sample and subvolume, expressed in decibels (dB).  

\begin{table}[htb!]
\centering
\resizebox{\linewidth}{!}{%
\begin{tabular}{|cc|cccc|c|}
    \toprule
     \midrule
    \textbf{Sample}& \textbf{Beamformer}& \textbf{Subvol. 1} & \textbf{Subvol. 2} & \textbf{Subvol. 3} & \textbf{Subvol. 4} & \textbf{Mean(SD)}\\
     \midrule
     \multirow{2}{*}{Sample 1}& DAS BF&7.2&11.6&8.1&8.8&8.9(1.9)\\
                             &Specular BF&14.8&13.5&9.6&10.2&12.0(2.5)\\
                             \multicolumn{2}{|r|}{$C_{EI}^{sp}-C_{EI}^{das}$}&7.6&1.9&1.5&1.4&3.1(3.0)\\
                             \midrule
     \multirow{2}{*}{Sample 2}& DAS BF&5.9&3.9&3.3&4.6&4.4(1.1)\\
                             &Specular BF&10.2&6.4&7.5&7.6&7.9(1.6)\\
                             \multicolumn{2}{|r|}{$C_{EI}^{sp}-C_{EI}^{das}$}&4.3&2.5&4.2&3.0&3.5(0.9)\\
                             \midrule
     \multirow{2}{*}{Sample 3}& DAS BF&-0.7&-1.4&-1.1&-1.5&-1.2(0.4)\\
                             &Specular BF&-0.8&-1.0&1.4&0.4&-0.0(1.1)\\
                             \multicolumn{2}{|r|}{$C_{EI}^{sp}-C_{EI}^{das}$}&-0.1&0.4&2.5&1.5&1.1(1.2)\\
                             \midrule
     \multirow{2}{*}{Sample 5}& DAS BF&-&-&-&-&-\\ 
     &Specular BF&-&-&-&-&-\\ 
                             \multicolumn{2}{|r|}{$C_{EI}^{sp}-C_{EI}^{das}$}&-&-&-&-&-\\
                             \midrule
     \bottomrule
\end{tabular}
}
\caption{Summary of  endosteal visibility metric $C_{EI}$ computed for each sample and each subvolume expressed in dB}
\label{tab_chap6:summary_cei}
\end{table}

For sample 1, the endosteal interface is clearly visible across all subvolumes with DAS beamforming ($C_{EI}$ ranging from 7.2 to 11.6 dB). Specular beamforming significantly enhances endosteal interface visibility, with $C_{EI}$ values ranging from 10.2 to 14.8 dB, corresponding to an increase of 1.4 to 7.6 dB. Compared to DAS images, specular beamforming images exhibit reduced intra-cortical speckle and improved endosteal interface intensity. The specularity map highlights high specularities at the periosteal and endosteal interfaces ($C_{EI} = 7$ dB) and low specularities within the cortex. The local orientation of the interface is shown in the specular orientation map.  

For sample 2, the endosteal interface remains well-defined across all subvolumes with DAS beamforming ($C_{EI}$ ranging from 3.3 to 5.9 dB). Specular beamforming further enhances visibility, with $C_{EI}$ ranging from 6.4 to 10.2 dB, representing an increase of 2.5 to 4.3 dB.

In high-porosity samples (samples 3 and 5), the endosteal interface does not exhibit significantly higher visibility than intra-cortical speckle. For sample 3, small negative $C_{EI}$ values are observed in both DAS and specular beamformed images, indicating that the endosteal interface is barely distinguishable from cortical speckle. However, in other subvolumes, specular beamforming enhances endosteal interface intensity by approximately 0.4 to 2.5 dB compared to DAS beamforming. The specularity map reveals moderate specularity values at the endosteal interface and within the cortex.  

For the most porous and heterogeneous sample (sample 5), neither DAS nor specular BF successfully reveals the endosteal interface. The cortical image appears too thin, with cortical thickness below the bone wavelength, making it impossible to compute the endosteal interface visibility metric ($C_{EI}$). However, reconstructed ultrasound images show that global speckle is significantly reduced with specular BF compared to DAS BF. The specularity map shows high specularities only at the periosteal interface, with low specularities elsewhere.  

In summary, specular beamforming reduces intra-cortical speckle across all samples and enhances the brightness of the endosteal interface in homogeneous samples (samples 1, 2, and 3).

\subsection{In-vivo results}
In Figure~\ref{fig_chap6:beamformed_invivo}, a comparison between DAS beamforming and Specular beamforming for \invivo~data is presented. The reconstructed ultrasound images obtained with each beamforming algorithm are displayed for subjects 1 to 10 for only one representative measure per subject.

\setlength{\tabcolsep}{1pt}
\afterpage{
\begin{landscape}
\begin{figure}[htb!]
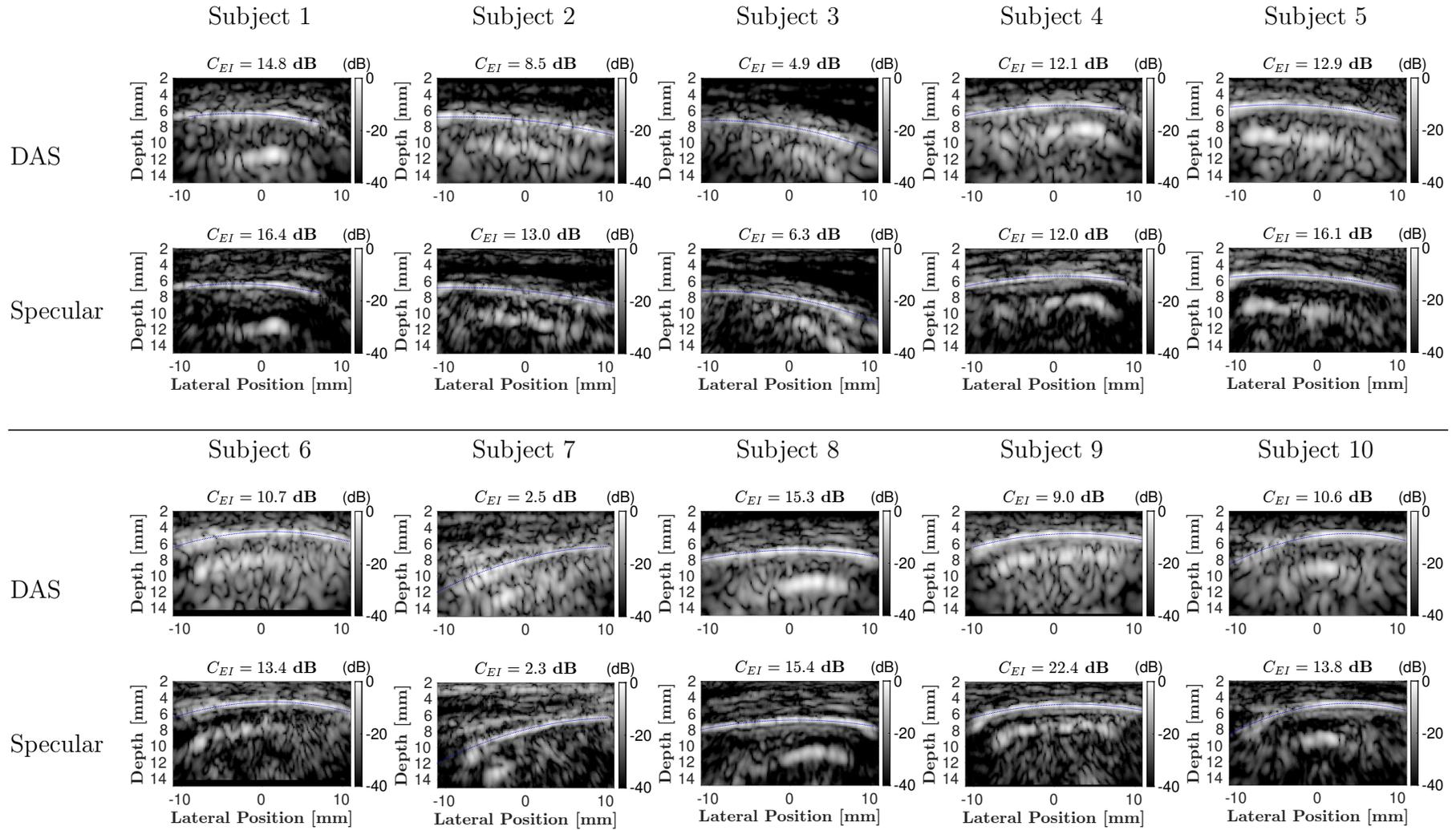

    \centering
    \renewcommand{\arraystretch}{0} 
    \begin{tabular}{m{.08\linewidth}m{.18\linewidth}m{.18\linewidth}m{.18\linewidth}m{.18\linewidth}m{.18\linewidth}}
    & \centerline{Subject 1} & \centerline{Subject 2} & \centerline{Subject 3} & \centerline{Subject 4} & \centerline{Subject 5}\\
    \end{tabular}
    \renewcommand{\arraystretch}{5} 
    \begin{tabular}{m{.08\linewidth}m{.18\linewidth}m{.18\linewidth}m{.18\linewidth}m{.18\linewidth}m{.18\linewidth}}
    DAS &\multirow{2}{*}{\begin{subfigure}{\linewidth}
        \includegraphics[width=\linewidth]{\imageProxiThird{09}{\stepPT}}
    \end{subfigure}}&
    \multirow{2}{*}{\begin{subfigure}{\linewidth}
        \includegraphics[width=\linewidth]{\imageProxiThird{10}{\stepPT}}
    \end{subfigure}}&
    \multirow{2}{*}{\begin{subfigure}{\linewidth}
        \includegraphics[width=\linewidth]{\imageProxiThird{11}{\stepPT}}
    \end{subfigure}}&
    \multirow{2}{*}{\begin{subfigure}{\linewidth}
        \includegraphics[width=\linewidth]{\imageProxiThird{13}{\stepPT}}
    \end{subfigure}}&
    \multirow{2}{*}{\begin{subfigure}{\linewidth}
        \includegraphics[width=\linewidth]{\imageProxiThird{14}{\stepPT}}
    \end{subfigure}}\\
    Specular & & & & &\\[1cm]
    \midrule
    \end{tabular}
    \renewcommand{\arraystretch}{0} 
    \begin{tabular}{m{.08\linewidth}m{.18\linewidth}m{.18\linewidth}m{.18\linewidth}m{.18\linewidth}m{.18\linewidth}}
    & \centerline{Subject 6} & \centerline{Subject 7} & \centerline{Subject 8} & \centerline{Subject 9} & \centerline{Subject 10}\\
    \end{tabular}
    \renewcommand{\arraystretch}{5} 
    \begin{tabular}{m{.08\linewidth}m{.18\linewidth}m{.18\linewidth}m{.18\linewidth}m{.18\linewidth}m{.18\linewidth}}
    DAS &\multirow{2}{*}{\begin{subfigure}{\linewidth}
        \includegraphics[width=\linewidth]{\imageProxiThird{15}{\stepPT}}
    \end{subfigure}}&
    \multirow{2}{*}{\begin{subfigure}{\linewidth}
        \includegraphics[width=\linewidth]{\imageProxiThird{16}{\stepPT}}
    \end{subfigure}}&
    \multirow{2}{*}{\begin{subfigure}{\linewidth}
        \includegraphics[width=\linewidth]{\imageProxiThird{17}{\stepPT}}
    \end{subfigure}}&
    \multirow{2}{*}{\begin{subfigure}{\linewidth}
        \includegraphics[width=\linewidth]{\imageProxiThird{18}{\stepPT}}
    \end{subfigure}}&
    \multirow{2}{*}{\begin{subfigure}{\linewidth}
        \includegraphics[width=\linewidth]{\imageProxiThird{19}{\stepPT}}
    \end{subfigure}}
    \\
    Specular & & & & &\\[1cm]
    \end{tabular}

    \caption{Comparison of DAS beamforming and specular beamforming for \invivo~ultrasound imaging. Ultrasound images for subjects 1 to 10 are shown, with the blue dashed line representing the parabolic fit of the periosteal interface segmentation.}
    \label{fig_chap6:beamformed_invivo}
\end{figure}
\end{landscape}}

DAS beamforming consistently yields images with good quality and good visibility of the endosteal interface across all subjects. However, within and after the cortex, there remains a significant presence of intra-cortical speckle.

On the other hand, Specular beamforming maintains good visibility of the endosteal interface while effectively reducing intra-cortical speckle compared to DAS beamforming. Over the 10 subjects, the DAS algorithm yields a mean and standard deviation ($C_{EI}$) of 10.1 dB (±3.9), while the Specular BF algorithm results in 13.0 dB (±5.3). This corresponds to an average increase of 3~dB.

Furthermore, an important effect of specular beamforming in the \invivo~data, not always visible in the \exvivo~images, is the reduction in speckle from the medullary cavity (after the endosteal interface) as well. This reduction in speckle both before and after the endosteal surface, while preserving a bright endosteal surface, enhances the overall contrast and image quality. However, it is worth noting that this improvement is not reflected in $C_{EI}$.

Despite the visual improvement observed with Specular beamforming, the metric of interface visibility ($C_{EI}$) may not accurately quantify this enhancement. There is significant variability in the degree of enhancement in $C_{EI}$ observed from one subject to another, ranging from very high (e.g., 13 dB for subject 9) to very low (e.g. less than 1 dB for subjects 4, 7, and 8), despite the visually improved images obtained with Specular beamforming for these subjects. This suggests that $C_{EI}$ might not be the most suitable metric for assessing the endosteal visibility, and alternative metrics may need to be considered to better evaluate the performance of Specular beamforming in \invivo~ultrasound imaging.

\afterpage{
\begin{landscape}

\begin{figure}[htb!]
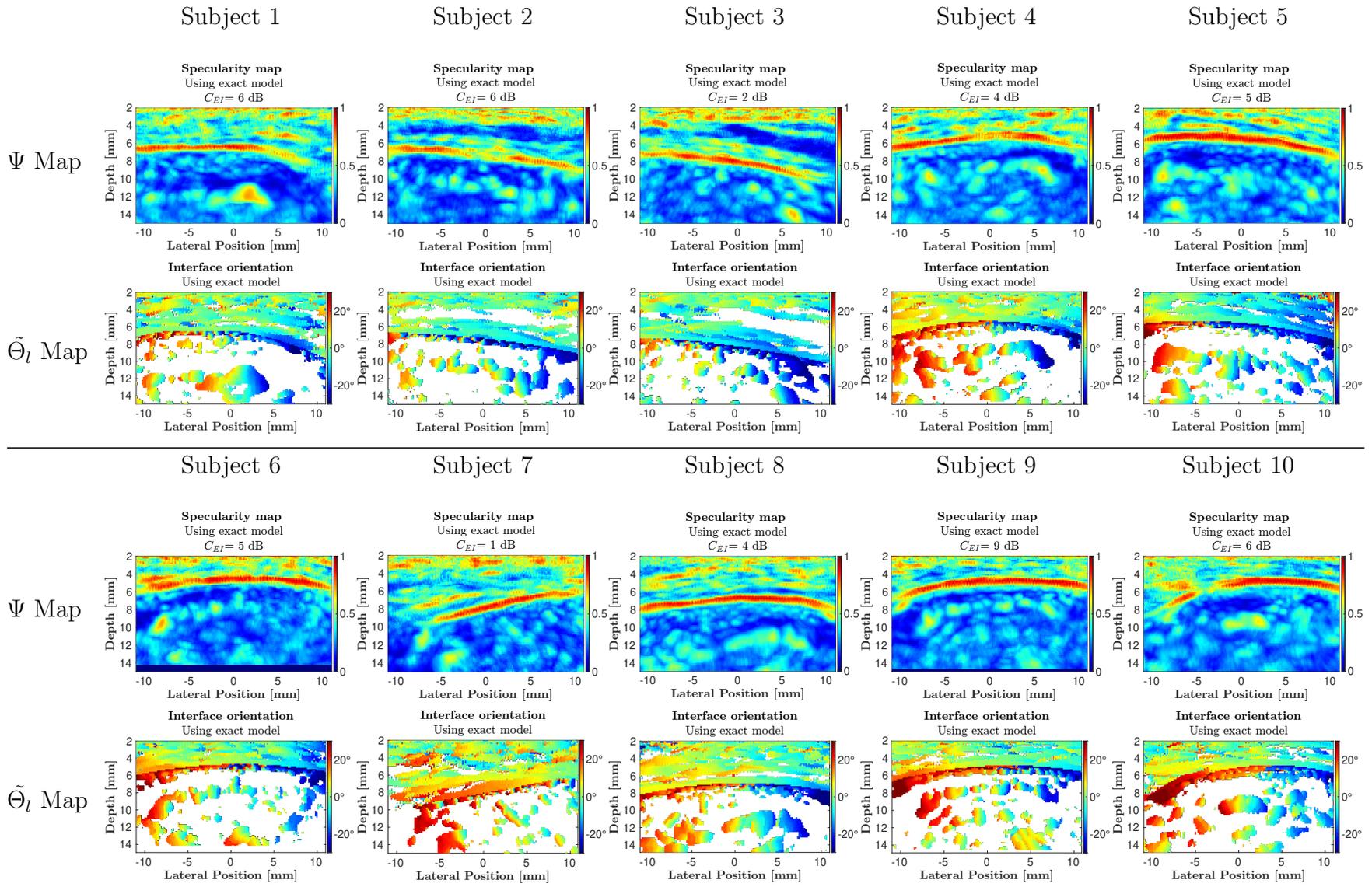

\setlength{\tabcolsep}{0pt}        
    \centering
    \renewcommand{\arraystretch}{1} 
    \begin{tabular}{m{.07\linewidth}m{.18\linewidth}m{.18\linewidth}m{.18\linewidth}m{.18\linewidth}m{.18\linewidth}}
    & \centerline{Subject 1} & \centerline{Subject 2} & \centerline{Subject 3} & \centerline{Subject 4}& \centerline{Subject 5}\\
    $\Psi$ Map &\begin{subfigure}{\linewidth}
        \includegraphics[width=\linewidth]{\imageSpeculaityProxiThird{09}{\stepPT}}
    \end{subfigure}&
    \begin{subfigure}{\linewidth}
        \includegraphics[width=\linewidth]{\imageSpeculaityProxiThird{10}{\stepPT}}
    \end{subfigure}&
    \begin{subfigure}{\linewidth}
        \includegraphics[width=\linewidth]{\imageSpeculaityProxiThird{11}{\stepPT}}
    \end{subfigure}&
    \begin{subfigure}{\linewidth}
        \includegraphics[width=\linewidth]{\imageSpeculaityProxiThird{13}{\stepPT}}
    \end{subfigure}&
    \begin{subfigure}{\linewidth}
        \includegraphics[width=\linewidth]{\imageSpeculaityProxiThird{14}{\stepPT}}
    \end{subfigure}\\
    $\Tilde{\Theta_l}$ Map & \begin{subfigure}{\linewidth}
        \includegraphics[width=\linewidth]{\imageOrientationProxiThird{09}{\stepPT}}
    \end{subfigure}&
    \begin{subfigure}{\linewidth}
        \includegraphics[width=\linewidth]{\imageOrientationProxiThird{10}{\stepPT}}
    \end{subfigure}&
    \begin{subfigure}{\linewidth}
        \includegraphics[width=\linewidth]{\imageOrientationProxiThird{11}{\stepPT}}
    \end{subfigure}&
    \begin{subfigure}{\linewidth}
        \includegraphics[width=\linewidth]{\imageOrientationProxiThird{13}{\stepPT}}
    \end{subfigure}&
    \begin{subfigure}{\linewidth}
        \includegraphics[width=\linewidth]{\imageOrientationProxiThird{14}{\stepPT}}
    \end{subfigure}\\
    \midrule
    & \centerline{Subject 6} & \centerline{Subject 7} & \centerline{Subject 8} & \centerline{Subject 9} & \centerline{Subject 10}\\
    $\Psi$ Map &\begin{subfigure}{\linewidth}
        \includegraphics[width=\linewidth]{\imageSpeculaityProxiThird{15}{\stepPT}}
    \end{subfigure}&
    \begin{subfigure}{\linewidth}
        \includegraphics[width=\linewidth]{\imageSpeculaityProxiThird{16}{\stepPT}}
    \end{subfigure}&
    \begin{subfigure}{\linewidth}
        \includegraphics[width=\linewidth]{\imageSpeculaityProxiThird{17}{\stepPT}}
    \end{subfigure}&
    \begin{subfigure}{\linewidth}
        \includegraphics[width=\linewidth]{\imageSpeculaityProxiThird{18}{\stepPT}}
    \end{subfigure}&
    \begin{subfigure}{\linewidth}
        \includegraphics[width=\linewidth]{\imageSpeculaityProxiThird{19}{\stepPT}}
    \end{subfigure}\\
    $\Tilde{\Theta_l}$ Map& \begin{subfigure}{\linewidth}
        \includegraphics[width=\linewidth]{\imageOrientationProxiThird{15}{\stepPT}}
    \end{subfigure}&
    \begin{subfigure}{\linewidth}
        \includegraphics[width=\linewidth]{\imageOrientationProxiThird{16}{\stepPT}}
    \end{subfigure}&
    \begin{subfigure}{\linewidth}
        \includegraphics[width=\linewidth]{\imageOrientationProxiThird{17}{\stepPT}}
    \end{subfigure}&
    \begin{subfigure}{\linewidth}
        \includegraphics[width=\linewidth]{\imageOrientationProxiThird{18}{\stepPT}}
    \end{subfigure}&
    \begin{subfigure}{\linewidth}
        \includegraphics[width=\linewidth]{\imageOrientationProxiThird{19}{\stepPT}}
    \end{subfigure}\\
    \midrule
    \end{tabular}

    \caption{Specularities and corresponding specular orientation of \invivo~ultrasound data. Images are shown for subject 1 to 10.}
    \label{fig_chap6:specularity_invivo}
\end{figure}
\end{landscape}}

Figure~\ref{fig_chap6:specularity_invivo} presents the specularity ($\Psi$) and the local orientation maps ($\Tilde{\Theta_l}$) of pixels with specularities greater than 0.4 for all subjects. For each subject, the top row displays the specularity maps, while the bottom row shows the corresponding local orientation maps.

The specularity of the periosteal interface appears consistently high across all subjects, with values exceeding 0.7, indicating robust specular reflections at this interface. Moderately high specularity values, exceeding 0.5, are observed at the endosteal interface, suggesting a notable specular reflection but to a lesser extent than the periosteal interface. Within the cortex, pixels with moderate specularity (greater than 0.5) are present but lack significant connectivity, implying that these areas may not represent genuine specular structures within the cortex.

The estimated local orientations align well with the observed geometry in the ultrasound images, indicating consistency between the specularity maps and the underlying bone structure.

In summary, the analysis of specularity combined with local orientation maps tells us that the bright interfaces present in the ultrasound images correspond to specular interface.

\section{Conclusion \& Discussion}
    In this study, we investigated the efficacy of considering the endosteal surface as a specular interface in enhancing its visibility in the ultrasound images. We applied specular beamforming to both \exvivo~and \invivo~human cortical bone. The images obtained are compared to those obtained using the beamforming method of Delay-And-Sum. A visibility metric for the interface was provided to quantify the improvements achieved.

    Specular beamforming consistently enhances the visibility of the endosteal interface compared to DAS beamforming, as observed in both \exvivo~and \invivo~experiments. This enhancement is evidenced by higher endosteal visibility metric ($C_{EI}$) values obtained with specular beamforming. Notably, specular BF images exhibit significantly reduced speckle surrounding the endosteal interface while maintaining a bright interface, resulting in clearer images of bone geometry. The reduction of speckle from the marrow (medullary cavity) is particularly pronounced in the \invivo~ultrasound images compared to the \exvivo~images.

    Despite improvements in image quality with specular beamforming, $C_{EI}$ may not accurately quantify these improvements. $C_{EI}$ does not consider the reduction of the speckle in the medullary cavity and in some \invivo~cases, despite visual improvements in images obtained with specular beamforming, $C_{EI}$ values may be lower.

\paragraph{Model curvature} 
In this study, we assumed planar specular interfaces in the computation of the model of specular transform. While this assumption simplifies the computational process, it becomes evident from the micro-CT images that the curvature of the samples is significant. Tis results in an underestimation of the specularity of the interface and inaccuracies in the specular orientation estimation. Therefore, although the assumption of a planar specular interface may be appropriate for certain applications where curvature effects are minimal or negligible, it is essential to acknowledge its limitations.
    \paragraph{Perspectives} 
    Overall, the application of specular beamforming demonstrates promise for enhancing the quality of ultrasound imaging in bone tissue, offering improved visualization of endosteal interface and potentially aiding in improving the estimation of cortical thickness. However, further research is needed to optimize parameters, refine models, and develop more sensitive metrics to fully harness the benefits of specular beamforming in bone ultrasound imaging.

    A combined analysis of specularity and local orientation maps can offer more information into the characteristics of the endosteal interface. By leveraging both outputs, it becomes possible to derive a novel quantity that assesses the roughness of the endosteal interface. This approach may provide a more comprehensive understanding of the interface's properties and enhance our ability to characterize bone microstructure accurately.
    
    The wave speeds used for computing delays are obtained through an autofocus approach using a DAS algorithm applied by Renaud et al. to bone. However, employing this autofocus approach with a specular algorithm may lead to a more suitable and accurate estimation of the propagating wave speed. Additionally, the segmentation of periosteal and endosteal interfaces currently utilizes Dijkstra's algorithm on the DAS image. Considering the specific characteristics of specular beamforming, applying Dijkstra's algorithm directly to the specular BF image or to the specularity map may offer improved segmentation results. 
    
    The \invivo~feasibility of real-time specular beamforming requires code optimization. Obtaining a DAS image is approximately 100 timesfaster than obtaining an specular BF with the exact model. The majority of the time for specular beamforming is dedicated to computing the specular model. Simplifying this model could lead to faster computation without significantly compromising image quality. Currently, the specular BF algorithm is fully implemented using MATLAB 2023. 

    The scope of this study is limited to analyzing the radial direction of bone (transverse) for both \invivo~and \exvivo~data. While it could be extended to include the longitudinal configuration, caution must be taken as the axial direction corresponds to the direction of the pores, potentially leading to specular structures inside the cortex.

\clearpage
\bibliographystyle{unsrt}
\bibliography{biblio_specular}
\end{document}